\documentclass[12pt]{article}
\newcommand{\filename}{2LMM\_HP\_split\_v5}
\usepackage{graphicx}
\usepackage{authblk}
\usepackage{amsfonts}
\usepackage{amssymb}
\usepackage{amsmath}
\usepackage{amsthm}
\usepackage{framed}

\usepackage{enumitem}
\usepackage{geometry}
\usepackage{url}
\usepackage{booktabs}

\usepackage{multirow}

\usepackage{mdframed}
\usepackage{mathtools}
\usepackage{algorithm}  
\usepackage{algorithmic}

\newcommand{\Ctrain}{\mathcal{C}_\mathrm{train}}  
\newcommand{\Ctest}{\mathcal{C}_\mathrm{test}}

\newcommand{\lf}{\mathrm{lf}}

\newcommand{\eledeg}{\mathrm{eledeg}}

\newcommand{\ttH}{{\tt H}}  
\newcommand{\ttC}{{\tt C}}  
\newcommand{\ttO}{{\tt O}}  
\newcommand{\ttN}{{\tt N}}  
\newcommand{\ttP}{{\tt P}}  
\newcommand{\ttF}{{\tt F}}  
\newcommand{\ttCl}{{\tt Cl}}  
\newcommand{\ttS}{{\tt S}}  
  
\newcommand{\ttPb}{{\tt Pb}}  
\newcommand{\ttBr}{{\tt Br}}

\newcommand{\oH}{\overline{{\tt H}}}

\newcommand{\C}{\mathbb{C}}  
\newcommand{\Co}{\mathbb{C}}

\newcommand{\anC}{\langle \mathbb{C} \rangle}  
\newcommand{\anpsi}{\langle \psi \rangle}  

\newcommand{\VH}{V_{\tt H}}

\newcommand{\R}{\mathbb{R}} 
 
\newcommand{\RK}{\mathbb{R}^K}

\newcommand{\dcp}{\mathrm{dcp}}

\newcommand{\Vleaf}{V_\mathrm{leaf}} 
\newcommand{\Eleaf}{E_\mathrm{leaf}} 

\newcommand{\sint}{\sigma_\mathrm{int}} 
\newcommand{\sce}{\sigma_\mathrm{ce}}

\newcommand{\Ez}{E_{(0/1)}}
\newcommand{\Ew}{E_{(\geq 1)}}
\newcommand{\Et}{E_{(\geq 2)}}
\newcommand{\Eew}{E_{(=1)}}

\newcommand{\Gac}{\Gamma_\mathrm{ac}}

\newcommand{\w}{w}
\newcommand{\x}{x}

\newcommand{\ta}{{\tt a}}
\newcommand{\tb}{{\tt b}}
\newcommand{\Ldg}{\Lambda_{\mathrm{dg}}}

\newcommand{\fc}{\mathrm{fc}}

\newcommand{\val}{\mathrm{val}}

\newcommand{\inte}{\mathrm{int}}

\newcommand{\nint}{\mathrm{n}^\mathrm{int}}

\newcommand{\h}{\mathrm{ht}}

\newcommand{\cs}{\mathrm{cs}}
\newcommand{\ch}{\mathrm{ch}}

\newcommand{\dg}{\mathrm{dg}}

\newcommand{\na}{\mathrm{na}}

\newcommand{\acC}{\mathrm{ac}_\mathrm{C}}

\newcommand{\ns}{\mathrm{ns}}

\newcommand{\ec}{\mathrm{ec}}
\newcommand{\ac}{\mathrm{ac}}

\newcommand{\bl}{\mathrm{bl}}

\newcommand{\bd}{\mathrm{bd}}

\newcommand{\UB}{\mathrm{UB}}
\newcommand{\LB}{\mathrm{LB}}

\newcommand{\ex}{\mathrm{ex}}

\newcommand{\GC}{G_\mathrm{C}}

\newcommand{\VC}{V_\mathrm{C}}

\newcommand{\EC}{E_\mathrm{C}}

\begin{document} 

\begin{center}
   {\Large\bf 
    Molecular Design Based on 
    Integer Programming and Splitting Data Sets by Hyperplanes}
   \\ 
\end{center}

\begin{center} 
Jianshen Zhu$^1$, 
Naveed Ahmed Azam$^{2,*}$, 
Kazuya Haraguchi$^1$, 
Liang Zhao$^3$, 
Hiroshi Nagamochi$^1$ 
 and  
 Tatsuya Akutsu$^4$ 
\end{center} 
%
%
{\small 
 $^1$  Department of Applied Mathematics and Physics, Kyoto University, 
 Kyoto 606-8501, Japan\\
$^2$ Department of Mathematics, Quaid-i-Azam University, Islamabad 45320, Pakistan\\
$^3$   Graduate School of Advanced Integrated Studies in Human Survavibility
     (Shishu-Kan),   Kyoto University, Kyoto 606-8306, Japan \\
$^4$   Bioinformatics Center,  Institute for Chemical Research, 
  Kyoto University, Uji 611-0011, Japan \\
$^*$ Corresponding author\\
}

\begin{quote}  
{\bf Abstract}\\  
A novel framework  for designing 
the molecular structure of chemical compounds
with a desired chemical property has recently been proposed. 
%
%
The framework infers a desired chemical graph
by  solving  a mixed integer linear program (MILP)
that simulates the computation process of
a feature function  defined by a two-layered model on chemical graphs
and a prediction function constructed by a machine learning method. 
To improve the learning performance of prediction functions in the framework, 
 we design a method that splits  a given data set $\mathcal{C}$  into two subsets $\mathcal{C}^{(i)},i=1,2$ by a hyperplane in a chemical space
so that  most compounds in the first (resp., second)
subset have observed values lower (resp., higher) than 
a threshold $\theta$.
We construct a prediction function $\psi$ to the data set $\mathcal{C}$
by combining  prediction functions $\psi_i,i=1,2$ each of which 
is constructed on $\mathcal{C}^{(i)}$ independently.  
The results of our computational experiments suggest that  
 the proposed method improved the learning performance 
 for several chemical properties to which a good prediction function 
 has been difficult to construct. 



\end{quote}

\section{Introduction}\label{sec:introduction}
 
\noindent {\bf Background~}
Among various application areas of bioinformatics and machine learning,
drug design is gathering interest \cite{Lo18,Tetko20}.
Accordingly, extensive studies have been done on computational analysis
of chemical structures.
There are two major topics in such studies:
prediction of the chemical property of a given chemical structure,
and design of a chemical structure having a desired chemical property.
These topics have also been extensively studied in the field of
chemoinformatics, where the former one is referred to as
\emph{quantitative structure activity relationship} (QSAR) \cite{CMF14}
and the latter as
\emph{inverse quantitative structure activity relationship}
(inverse QSAR) \cite{Miyao16,Ikebata17,Rupakheti15}.

For the prediction task,
statistical methods and machine learning methods have been extensively utilized
\cite{CMF14}.
In most of such studies, there are two phases:
learning phase and prediction phase.
In the learning phase,
a prediction function is derived from
training data consisting of pairs of chemical structures and their activities
(or properties),
where each chemical structure is given as an undirected graph (called
a chemical graph) and then is transformed into a vector of real numbers
called \emph{features} or \emph{descriptors}.
In the prediction phase, the prediction function derived as above
is simply applied to the feature vector obtained from a given chemical graph.
To derive a prediction function,
regression-based methods have been utilized in traditional QSAR studies
\cite{CMF14},
where machine learning-based methods, including
artificial neural network (ANN)-based methods \cite{Ghasemi18,kim22},
have recently been extensively utilized.
It is to be noted that when using graph convolutional networks (GCNs)~\cite{Kipf16},
chemical graphs can be directly handled
and thus transformation to feature vectors is not necessarily required.

For the design task,
prediction functions are also utilized and are usually derived from
existing data as in the above.
Then, chemical structures are inferred from
given chemical activities through a prediction function
\cite{Miyao16,Ikebata17,Rupakheti15}
where additional constraints may be imposed to restrict
the possible structures.
In the traditional approach,
this inference task consists of two phases,
(i) derivation of feature vectors from given chemical activities
using the inverse of the prediction function,
and
(ii) reconstruction of chemical structures from given feature vectors,
where these two phases are often mixed.
For (i),
some optimization methods and/or sampling methods are usually
employed.
For (ii),
some enumeration methods are often applied.
However, both are inverse problems and are computationally difficult.
For example, it is known that
the number of possible chemical graphs is huge~\cite{BMG96} and
inference of a chemical graph from a given feature vector is
NP-hard in general~\cite{AFJS12}.
Therefore,
most existing methods employ heuristic methods for both (i) and (ii),
and thus do not guarantee optimal or exact solutions.

Recently, different approaches have been proposed for the design task
using ANNs.
One of the attractive points of ANNs is that generative models are available,
which include autoencoders and generative adversarial networks.
Furthermore, as mentioned before, chemical structures can be directly
handled by using GCNs~\cite{Kipf16}.
By combining these techniques, it might be possible to design novel
chemical structures without solving the inverse problems \cite{xiong22}.
Indeed, extensive studies have recently been done using various ANN models,
which include
variational autoencoders~\cite{Gomez18}, 
grammar variational autoencoders~\cite{Kusner17},
generative adversarial networks~\cite{DeCao18},
recurrent neural networks~\cite{Segler18,Yang17}, 
and invertible flow models~\cite{Madhawa19,Shi20}.
However,
these are heuristic methods (although based on some statistical models)
and thus do not guarantee optimality or exactness of the solutions.

\begin{figure}[!ht]  \begin{center}
\includegraphics[width=.77\columnwidth]{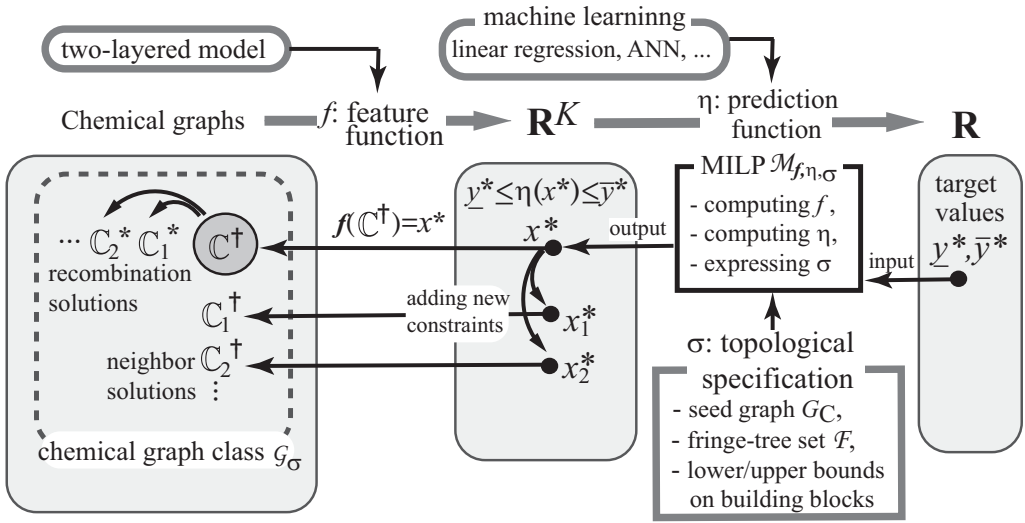}
\end{center} \caption{An illustration of inferring
desired chemical graphs $\Co\in \mathcal{G}_\sigma$ with
$\underline{y^*}\leq \eta(f(\Co))\leq \overline{y^*}$.   } 
\label{fig:framework}  \end{figure}    

\smallskip
\noindent {\bf Framework~}
A novel  framework for inferring chemical graphs has been developed
 \cite{SZAHZNA21,ZAHZNA21,poly_ICZAHZNA22,gridAZHZNA21} based on an idea of formulating
 as a mixed integer linear programming (MILP), 
 the computation process of a prediction function constructed
by a machine learning method.
The unique point of this framework is that
once a prediction function is formulated,
the inverse problem can be solved exactly by applying an MILP solver.
It consists of two main phases: the first phase 
 constructs a prediction function $\eta$ for a chemical property and the second phase 
infers a chemical graph with a target value of the property based on the function $\eta$. 
For a chemical property $\pi$, let $\mathcal{C}_{\pi}$ be a data set  of chemical graphs  
such that the observed value $a(\Co)$ of property $\pi$ 
 for every  chemical graph $\Co \in \mathcal{C}_{\pi}$ is available.
 In the first phase, we introduce a feature function $f: \mathcal{G}\to \mathbb{R}^K$ 
for a positive integer $K$,
where the descriptors of
a chemical graph are defined based on local graph  structures in a special way
called a {\em two-layered model}. 
We then construct a prediction function $\eta$
by a machine learning method such as linear regression, decision tree and an ANN
so that the output $y=\eta(x)\in \mathbb{R}$ of    
the feature vector  $x=f(\Co)\in \mathbb{R}^K$ for each $\Co\in \mathcal{C}_{\pi}$
  serves as a predicted value to the real value $a(\Co)$.  
In the second phase of inferring a desired chemical graph,
we specify not only a target chemical value for property $\pi$ but also
an abstract structure for a chemical graph to be inferred.
The latter is described by a set of rules based on the two-layered model called 
a {\em topological specification} $\sigma$, and denote by $\mathcal{G}_\sigma$
 the set of all chemical graphs that satisfy the rules in $\sigma$. 
The users select topological specification  $\sigma$ and
 two reals $\underline{y}^*$ and $\overline{y}^*$ 
 as an interval  for a  target chemical value.
The task of the  second phase is 
to infer  chemical graphs $\Co^*\in \mathcal{G}_\sigma$
such that  
 $\underline{y}^*\leq \eta(f(\C^*))\leq \overline{y}^*$
(see Figure~\ref{fig:framework}  for an illustration).
For this, we formulate an MILP $\mathcal{M}_{f,\eta,\sigma}$
that represents 
(i) the  computation process of  $x:=f(\Co)$ from a chemical graph $\Co$
in the feature function $f$;
(ii) the  computation process of  $y:=\eta(x)$ from a vector $x\in\mathbb{R}^K$
in the prediction function $\eta$; and
(iii) the  constraint for  $\Co\in  \mathcal{G}_\sigma$.
Given an interval with $\underline{y}^*,\overline{y}^* \in \mathbb{R}$,
 we solve the MILP $\mathcal{M}_{f,\eta,\sigma}$
to find  a feature vector $x^*\in \mathbb{R}^K$
 and a chemical graph $\Co^{\dagger}\in  \mathcal{G}_\sigma$ 
 such that $f(\Co^\dagger)=x^*$ and  
$\underline{y}^*\leq \eta(x^*) \leq \overline{y}^*$
(where if the MILP instance is infeasible
 then this suggests that $\mathcal{G}_\sigma$ 
does  not contain such a desired chemical graph). 
In the second phase, we next generate 
  some other desired chemical graphs based on the solution $\Co^{\dagger}$.
For this, the following two methods have been designed. 

 The first method  constructs 
   isomers of   $\C^\dagger$   without solving any new MILP.
In this method, we first decompose the chemical graph $\C^\dagger$ into 
a set of chemical acyclic graphs $T^\dagger_1,T^\dagger_2,\ldots,T^\dagger_q$,
and next construct  a set $\mathcal{T}_i$ of isomers $T^*_i$ of each tree $T^\dagger_i$
 such that $f(T^*_i)=f(T^\dagger_i)$ by a dynamic programming algorithm
 due to Azam~et~al.~\cite{AZSSSZNA20}.  
 Finally 
 we choose an isomer $T^*_i\in \mathcal{T}_i$ for each $i=1,2,\ldots,q$
 and  assemble them into an isomer $\C^*\in \mathcal{G}_\sigma$ of $\C^\dagger$
 such that $f(\C^*)=x^*=f(\C^\dagger)$.
  The first method generates such isomers $\C_1^*, \C_2^*, \ldots$ 
 which
 we call {\em  recombination solutions} of $\C^\dagger$.

The second method constructs   new solutions 
by solving the MILP $\mathcal{M}_{f,\eta,\sigma}$
with an additional set $\Theta$ of new linear constraints~\cite{gridAZHZNA21}.
We first prepare arbitrary $p_{\mathrm{dim}}$
 linear functions $\theta_j: \RK\to \R, j=1,2,\ldots,p_{\mathrm{dim}}$
and consider a neighbor of $\C^\dagger$ 
 defined by a set of chemical graphs $\C^*$ that satisfy linear constraints 
 $(k-0.5)\delta\leq |\theta_j(f(\C^*))-\theta_j(f(\C^\dagger))|\leq (k+0.5)\delta,  
 j=1,2,\ldots,p_{\mathrm{dim}}$ 
 for  a small real  $\delta>0$ and an integer $k\geq 1$.  
By changing the  integer $k$ systematically, we can search for new solutions
$\C^\dagger_1,\C^\dagger_2,  \ldots \in \mathcal{G}_\sigma$ 
of  MILP $\mathcal{M}_{f,\eta,\sigma}$
with constraint $\Theta$ such that the feature
vectors $x^*=f(\C^\dagger), x^*_1=f(\C^\dagger_1), x^*_2=f(\C^\dagger_2),\ldots$ 
are all slightly different.
We call these chemical graphs
 $\C^\dagger_1,\C^\dagger_2,  \ldots$ {\em neighbor solutions} of $\C^\dagger$.


The main reason why the framework can infer a chemical compound
with 50 non-hydrogen atoms is that  the descriptors of
a chemical graph are defined  on local graph structures in the two-layered model  and
thereby an MILP necessary to represent
a chemical graph can be formulated 
as a considerably compact form that is efficiently solvable by a standard solver.

\smallskip
\noindent {\bf Contribution~}
The descriptors   in the framework
  mainly consists of  the frequencies of local graph structures
based on the two-layered model
by which a chemical graph $\Co$ is regarded as a pair
of interior and exterior structures (see Section~\ref{sec:2LM} for details).
To derive a compact MILP formulation to infer a chemical graph,
it is important to use the current definition of descriptors.
However, there are some chemical properties for which
the performance of a prediction function constructed 
with the feature function $f$ remains rather low.
To improve the learning performance of prediction functions with the same two-layered model,
we propose a method of splitting a given data set with  a hyperplane in the feature space 
 into two subsets, where 
we construct a prediction function to each of the subsets independently before 
a prediction function $\eta$ to the original set is obtained by
combining these prediction functions (see Section~\ref{sec:LPs} for details).
Based on the same MILP  $\mathcal{M}_{f,\eta,\sigma}$ formulation 
proposed by  Zhu~et~al.~\cite{ZAHZNA21},
we implemented the framework to treat the newly proposed type of prediction.
From the results of our computational experiments on over some chemical properties
such as odor threshold~\cite{pubchem},
we observe that our new method of splitting data sets and combining prediction functions
improved the performance of a prediction function for these chemical properties.  
It is to be noted that extensive studies have been done on prediction problems
using hyperplanes 
since the development of support vector machines \cite{Cortes95}.
However, existing methods can only be applied to prediction problems.
The novel and unique point of our study is that
an efficient MILP formulation for chemical graphs is developed,
which makes the inverse problem (i.e., design problem) tractable.

The paper is organized as follows.  
Section~\ref{sec:preliminary} introduces some notions on graphs and 
 a modeling of chemical compounds. 
Section~\ref{sec:2LM} reviews  the two-layered model and
 descriptors defined by the model.
 Section~\ref{sec:prediction_functions} 
  reviews prediction functions constructed by linear regression.
Section~\ref{sec:LPs} proposes a method of splitting a data set by a hyperplane
and a linear programming formulation for finding such a hyperplane.
Section~\ref{sec:experiment} reports the results on  computational 
experiments conducted for 22 chemical properties 
such as autoignition temperature, flammable limits and odor threshold. 
Section~\ref{sec:conclude} makes some concluding remarks.   
Some technical details are given in Appendices:   
 Appendix~\ref{sec:descriptor}
    for  all descriptors in our feature function; 
 Appendix~\ref{sec:specification}
   for a full description of a topological specification; and 
Appendix~\ref{sec:test_instances}
  for the detail of test instances used in our computational experiment.
%

\section{Preliminary}\label{sec:preliminary}

This section  introduces some notions and terminologies on graphs,
  modeling of chemical compounds and our choice of descriptors. 
 
Let $\mathbb{R}$, $\mathbb{R}_+$, $\mathbb{Z}$  and $\mathbb{Z}_+$ 
denote the sets of reals,  non-negative reals, 
integers and non-negative integers, respectively.
For two integers $a$ and $b$, let $[a,b]$ denote the set of 
integers $i$ with $a\leq i\leq b$.
For a vector $x\in \R^p$, the $j$-th entry of $x$ is denoted by $x(j)$.

\bigskip\noindent
{\bf  Graph} 
Given a  graph $G$, let $V(G)$ and $E(G)$ denote the sets
of vertices and edges, respectively.     
For a subset $V'\subseteq V(G)$ (resp., $E'\subseteq E(G))$ of
a graph $G$, 
let $G-V'$ (resp., $G-E'$) denote the graph obtained from $G$
by removing the vertices in $V'$ (resp.,  the edges in $E'$),
where we remove all edges incident to a vertex in $V'$ to obtain $G-V'$. 
%
A path with two end-vertices $u$ and $v$ is called a {\em $u,v$-path}. 
 
 We define a {\em rooted} graph to be
 a graph with a  designated vertex, called a {\em root}. 
 %
%
 For a graph $G$ possibly with a root,  
 a {\em leaf-vertex} is defined to be a non-root vertex 
 with degree 1.
 Call  the edge $uv$ incident to a leaf vertex $v$ a {\em leaf-edge},
 and denote by $\Vleaf(G)$ and $\Eleaf(G)$
  the sets of leaf-vertices and leaf-edges  in $G$, respectively.
 For a graph  or a rooted graph $G$,
 we define graphs $G_i, i\in \mathbb{Z}_+$ obtained from $G$
 by removing the set of leaf-vertices $i$ times so that
\[ G_0:=G; ~~ G_{i+1}:=G_i - \Vleaf(G_i), \]
where we call a vertex $v$ a {\em tree vertex} if $v\in \Vleaf(G_i)$
for some $i\geq 0$. 
Define the {\em height} $\h(v)$ of each tree vertex $v\in \Vleaf(G_i)$
to be $i$; and 
$\h(v)$ of each non-tree vertex $v$ adjacent to a tree vertex 
to be $\h(u)+1$ for the maximum $\h(u)$ of a tree vertex $u$ adjacent to $v$,
where we do not define height of any non-tree vertex not adjacent to any tree vertex. 
We call a vertex $v$ with $\h(v)=k$ a {\em leaf $k$-branch}.
The {\em height} $\h(T)$ of a rooted tree $T$ is defined
to be the maximum of $\h(v)$ of a vertex $v\in V(T)$. 
 
\subsection{Modeling of Chemical Compounds}\label{sec:chemical_model}

We review a modeling of chemical compounds introduced 
by  Zhu~et~al.~\cite{ZAHZNA21}. 

To represent a chemical compound, 
we introduce a set  of   chemical elements such as 
  {\tt H} (hydrogen),  {\tt C} (carbon), {\tt O} (oxygen), {\tt N} (nitrogen)  and so on.
 To distinguish a chemical element $\ta$ with multiple valences such as {\tt S} (sulfur),
 we denote a chemical element $\ta$ with a valence $i$ by $\ta_{(i)}$,
 where we do not use such a suffix $(i)$ 
 for a chemical element $\ta$ with a unique valence. 
Let $\Lambda$ be a set of chemical elements $\ta_{(i)}$.
For example,  $\Lambda=\{\ttH,  \ttC, \ttO, \ttN, \ttP, \ttS_{(2)}, \ttS_{(4)}, \ttS_{(6)}\}$. 
Let $\val: \Lambda\to [1,6]$ be a valence function.
For example, $\val(\ttH)=1$, $\val(\ttC)=4$, $\val(\ttO)=2$, $\val(\ttP)=5$,
$\val(\ttS_{(2)})=2$, $\val(\ttS_{(4)})=4$ and $\val(\ttS_{(6)})=6$.
 For each  chemical element $\ta\in \Lambda$, 
let $\mathrm{mass}(\ta)$  denote the mass   of  $\ta$.

A chemical compound  is represented by a {\em chemical graph} defined to be
a tuple $\Co=(H,\alpha,\beta)$  of
  a simple, connected undirected graph $H$ and  
    functions   $\alpha:V(H)\to \Lambda$  and  $\beta: E(H)\to [1,3]$.
The set of atoms and the set of bonds in the compound 
are represented by the vertex set $V(H)$ and the edge set $E(H)$, respectively.
The chemical element assigned to a vertex $v\in V(H)$
is represented by $\alpha(v)$ and 
 the bond-multiplicity  between two adjacent vertices  $u,v\in V(H)$
is represented by $\beta(e)$ of the edge $e=uv\in E(H)$.
We say that two tuples $(H_i,\alpha_i,\beta_i), i=1,2$ are
{\em isomorphic} if they admit an isomorphism $\phi$,
i.e.,  a bijection $\phi: V(H_1)\to V(H_2)$
such that
 $uv\in E(H_1), \alpha_1(u)=\ta, \alpha_1(v)=\tb, \beta_1(uv)=m$
 $\leftrightarrow$  
 $\phi(u)\phi(v) \in E(H_2), \alpha_2(\phi(u))=\ta, 
 \alpha_2(\phi(v))=\tb, \beta_2(\phi(u)\phi(v))=m$. 
 When $H_i$ is rooted at a vertex $r_i,  i=1,2$,
these chemical graphs $(H_i,\alpha_i,\beta_i),  i=1,2$ are
{\em rooted-isomorphic} (r-isomorphic) if 
they admit  an isomorphism $\phi$ such that $\phi(r_1)=r_2$. 

 For a notational convenience, we  use
 a function $\beta_\Co: V(H)\to [0,12]$ 
 for a chemical graph $\Co=(H,\alpha,\beta)$
  such that $\beta_\Co(u)$ means the sum of bond-multiplicities
 of edges incident to a vertex $u$; i.e., 
\[ \beta_\Co(u) \triangleq \sum_{uv\in E(H) }\beta(uv) 
\mbox{ for each vertex $u\in V(H)$.}\]
For each vertex $u\in V(H)$, 
 define the {\em electron-degree} $\eledeg_\Co(u)$  to be 
\[  \eledeg_\Co(u) \triangleq  \beta_\Co(u) - \val(\alpha(u)). \]
For each  vertex $u\in V(H)$, let $\deg_\Co(v)$ denote 
the number of vertices adjacent to $u$ in $\Co$. 
  
  For a chemical   graph  $\Co=(H,\alpha,\beta)$, 
  let  $V_{\ta}(\Co)$, $\ta\in \Lambda$
   denote the set of vertices $v\in V(H)$ such that $\alpha(v)=\ta$ in $\Co$
  and define the {\em hydrogen-suppressed chemical graph} $\anC$ 
to be  the graph obtained from $H$ by
  removing all the vertices $v\in \VH(\Co)$.

\begin{figure}[h!] \begin{center}
\includegraphics[width=.80\columnwidth]{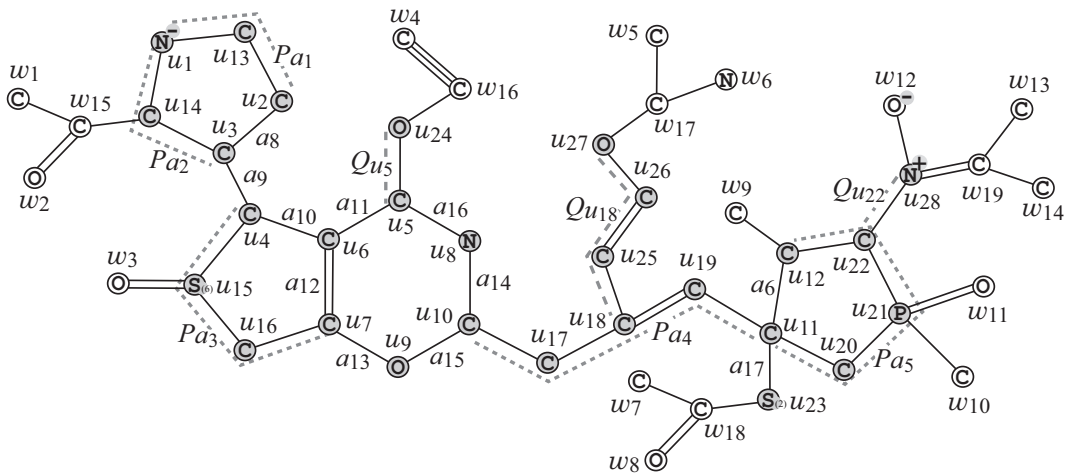}
\end{center} \caption{An illustration of  a hydrogen-suppressed chemical graph  
$\anC$ obtained from a chemical graph $\C$ 
by removing all the 
 hydrogens, 
where for  ${\rho}=2$,  
$V^\ex(\C)=\{w_i \mid i\in [1,19]\}$ and
$V^\inte(\C)=\{u_i \mid i\in [1,28]\}$.  
 }
\label{fig:example_chemical_graph} \end{figure}

\section{Two-layered Model}\label{sec:2LM}
This section reviews the two-layered model introduced  by 
 Shi~et~al.~\cite{SZAHZNA21}.
 
 Let  $\C=(H,\alpha,\beta)$ be a chemical graph
 and  ${\rho}\geq 1$ be an integer, which we call a {\em branch-parameter}.
 
  A {\em two-layered model} of $\C$ is a partition of
 the hydrogen-suppressed chemical graph $\anC$ into
 an ``interior'' and an ``exterior'' in the following way. 
 We call a vertex $v\in V(\anC)$
   (resp., an edge $e\in E(\anC))$ of   $\C$
   an {\em exterior-vertex} (resp.,    {\em exterior-edge}) if
    $\h(v)< {\rho}$ (resp., $e$ is incident to an  exterior-vertex)
and denote the sets of exterior-vertices and exterior-edges 
by $V^\ex(\C)$ and $E^\ex(\C)$, respectively
and denote  $V^\inte(\C)=V(\anC)\setminus  V^\ex(\C)$ and 
$E^\inte(\C)=E(\anC)\setminus E^\ex(\C)$, respectively.
We call a vertex in $V^\inte(\C)$ (resp.,   an edge in $E^\inte(\C)$) 
   an {\em interior-vertex} (resp.,    {\em interior-edge}). 
 The set  $E^\ex(\C)$ of  exterior-edges forms 
a collection of connected graphs each of which is
regarded as a rooted tree $T$ rooted at 
the vertex $v\in V(T)$ with the maximum $\h(v)$. 
Let $\mathcal{T}^\ex(\anC)$ denote 
the set of these chemical rooted trees in $\anC$. 
The {\em interior} $\C^\inte$ of $\C$ is defined to be the subgraph
 $(V^\inte(\C),E^\inte(\C))$ of $\anC$. 

Figure~\ref{fig:example_chemical_graph}
 illustrates an example of a hydrogen-suppressed chemical graph $\anC$.
For a branch-parameter ${\rho}=2$, 
the interior of  the chemical graph $\anC$ in Figure~\ref{fig:example_chemical_graph} 
is obtained by removing the set of vertices with degree 1 ${\rho}=2$ times; i.e., 
first remove  
the set  $V_1=\{w_1,w_2,\ldots,w_{14}\}$ of vertices of degree 1 in $\anC$ 
and then remove  the set
 $V_2=\{w_{15},w_{16},\ldots,w_{19}\}$ of vertices of degree 1 in $\anC-V_1$,
 where the removed vertices become the exterior-vertices of $\anC$.

\begin{figure}[h!] \begin{center}
\includegraphics[width=.84\columnwidth]{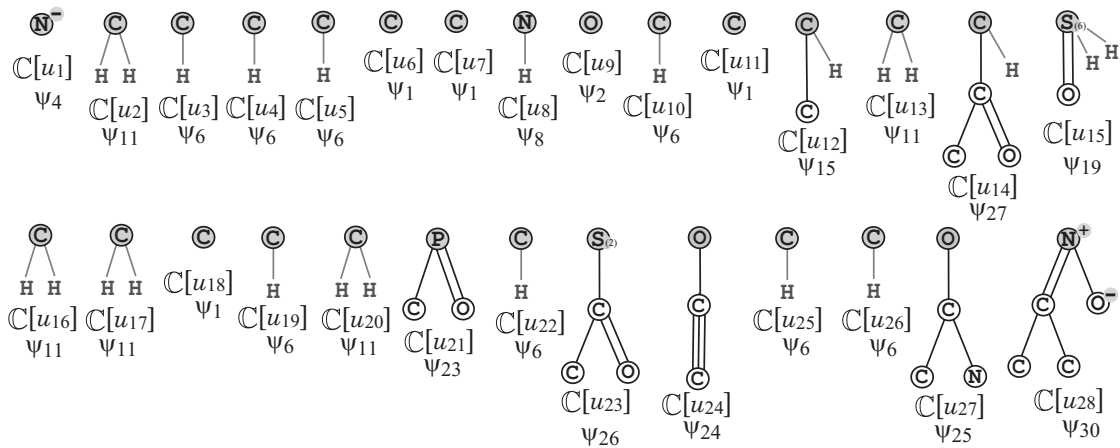}
\end{center} \caption{
The set $\mathcal{T}(\C)$ of  2-fringe-trees  $\C[u_i], i\in [1,28]$ of the example $\C$
in Figure~\ref{fig:example_chemical_graph}, 
where the root of each tree is depicted with a gray circle and
 the hydrogens attached to non-root vertices are omitted in the figure.  
 }
\label{fig:example_fringe-tree} \end{figure}

For each interior-vertex $u\in V^\inte(\C)$,
let $T_u\in \mathcal{T}^\ex(\anC)$ denote the chemical tree rooted at $u$
(where possibly $T_u$ consists of vertex $u$)
and 
define the {\em $\rho$-fringe-tree} $\C[u]$ to be  
the chemical rooted tree obtained from $T_u$ by putting back
 the hydrogens originally attached with $T_u$ in $\C$. 
Let $\mathcal{T}(\C)$ denote the set of $\rho$-fringe-trees 
$\C[u], u \in V^\inte(\C)$. 
Figure~\ref{fig:example_fringe-tree}  illustrates
the set  $\mathcal{T}(\C)=\{\C[u_i]\mid i\in [1,28]\}$ of the 2-fringe-trees 
  of the example $\C$
in Figure~\ref{fig:example_chemical_graph}. 

\smallskip
\noindent {\bf Feature Function~} 
 The feature of an  interior-edge $e=uv\in E^\inte(\C)$ 
 such that $\alpha(u)=\ta$, $\deg_{\anC}(u)=d$, 
 $\alpha(v)=\tb$, $\deg_{\anC}(v)=d'$  and $\beta(e)=m$  is represented by 
 a tuple $(\ta d, \tb d', m)$, which is called the {\em edge-configuration} 
  of the edge $e$, where 
  we call the tuple $(\ta, \tb, m)$ 
 the {\em adjacency-configuration} of the edge $e$. 
 

In the framework with the two-layered model,
the feature vector $f$ mainly consists of the frequency 
of edge-configurations of   the interior-edges  and
the frequency of chemical rooted trees among the set 
of  chemical rooted trees $\C[u]$ over all interior-vertices $u$. 
See Appendix~\ref{sec:descriptor} for  all these descriptors $x(1),x(2),\ldots,x(K_1)$,
which are called {\em linear descriptors}.
We denote by  $D_\pi^{(1)}:=\{x(k)\mid k\in[1,K_1]\}$ the set of descriptors
constructed over a data set for a property $\pi$. 
Zhu et al.~\cite{Q_ZACIHZNA22}\footnote{A full version of the article is available at \url{https://arxiv.org/abs/2209.13527}} introduced a quadratic term $x(i)x(j)$ (or $x(i)(1-x(j))$), $1\leq i\leq j\leq K_1$
as a new descriptor, where it is assumed that each $x(i)$ is normalized between 0 and 1.
This  term $x(i)x(j)$, $1\leq i\leq j\leq K_1$  (or $x(i)(1-x(j)), i,j\in [1,K_1]$)
is called a {\em quadratic descriptor} and is denoted by 
$D_\pi^{(2)}:=\{x(i)x(j)\mid 1\leq i\leq j\leq K_1\}\cup
 \{ x(i)(1-x(j))\mid  i,j\in [1,K_1]\}$ the set of  quadratic descriptors. 

To construct a prediction function, we use the union   $D_\pi^{(1)} \cup D_\pi^{(2)}$. 
 This set of descriptors is usually excessive in constructing a prediction function,  and 
 we reduce it to a smaller set of descriptors to construct a {\em feature function}
 $f: \RK\to \R$, where $K$ is the  number of resulting descriptors.
 We call  $\RK$ {\em  the feature space}.
 To reduce descriptors, we use the methods proposed 
 by Zhu et al.~\cite{Q_ZACIHZNA22}.

 
\smallskip
\noindent {\bf Topological Specification~}   
A topological specification $\sigma$ is described
as a set of the following rules:
\begin{enumerate}[nosep]
\item[(i)]
a {\em seed graph} $\GC$ as an  abstract form of  a target chemical graph $\C$;
\item[(ii)]
 a set $\mathcal{F}$ of chemical rooted trees  as candidates
 for a tree  $\C[u]$ rooted at each exterior-vertex $u$ in $\C$; 
and 
\item[(iii)]
lower and upper bounds on the number of components 
 in a target chemical graph such as  chemical elements, 
double/triple bonds and the interior-vertices in $\C$. 
\end{enumerate} 

\begin{figure}[h!] \begin{center}
\includegraphics[width=.98\columnwidth]{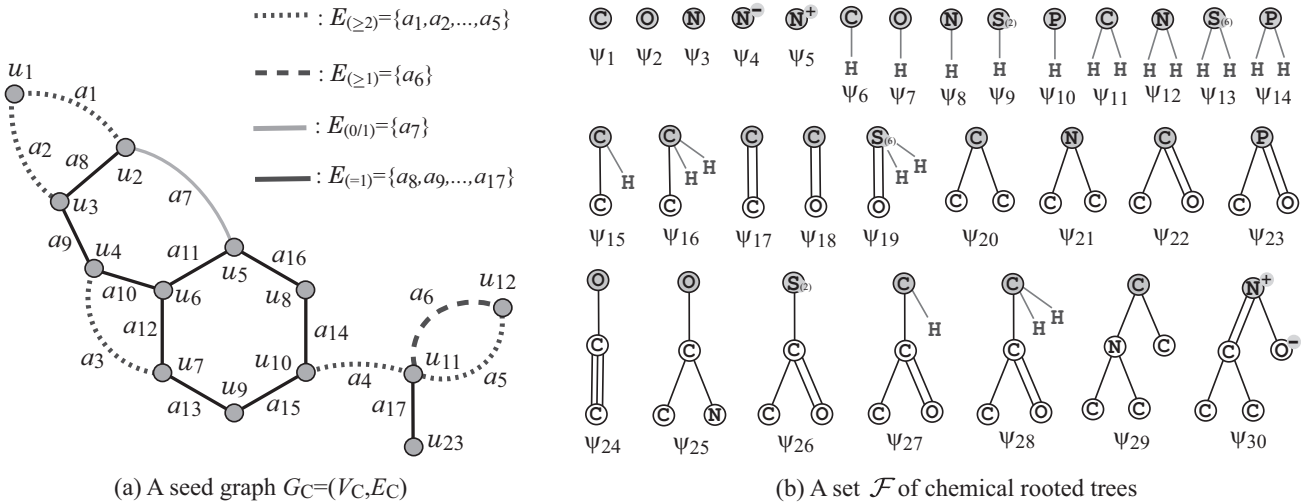}
\end{center} \caption{
(a) An illustration of a seed graph $\GC$, 
where the vertices in $\VC$ are depicted with gray circles,
the edges in $\Et$ are depicted with dotted lines,
the edges in $\Ew$ are depicted with dashed lines,
the edges in $\Ez$ are depicted with gray bold lines and  
the edges in $\Eew$ are depicted with black solid lines;
(b) A set $\mathcal{F}=\{\psi_1,\psi_2,\ldots,\psi_{30}\}\subseteq
\mathcal{F}(\mathcal{C}_\pi)$ of 30 chemical rooted trees
$\psi_i, i\in [1,30]$, where the root of each tree is depicted with a gray circle.
The hydrogens attached to non-root vertices are omitted in the figure.    }
\label{fig:specification_example_1} \end{figure}  

Figures~\ref{fig:specification_example_1}(a) and (b)
 illustrate  examples of  a  seed graph  $\GC$ and 
 a set $\mathcal{F}$ of chemical rooted trees, respectively. 
 Given a seed graph $\GC$, 
 the interior of   a target chemical graph $\C$ is constructed
 from $\GC$ by replacing some edges $a=uv$ 
 with paths $P_a$ between the end-vertices
 $u$ and $v$ and by attaching new paths $Q_v$ to some vertices $v$.  
%
For example, a chemical graph $\C$ 
in Figure~\ref{fig:example_chemical_graph} is constructed
from the seed  graph  $\GC$ in Figure~\ref{fig:specification_example_1}(a)
as follows.
\begin{enumerate}[nosep,  leftmargin=*]
\item[-]
First replace  five edges
 $a_1=u_1 u_{2},  a_2=u_1 u_{3},  a_3=u_4 u_{7}, a_4=u_{10}u_{11}$
and $a_5=u_{11}u_{12}$ in  $\GC$ 
 with new paths  
$P_{a_1}=(u_1,u_{13},u_{2})$, 
$P_{a_2}=(u_{1},u_{14},u_{3})$,
$P_{a_3}=(u_{4},u_{15},u_{16},u_{7})$, 
 $P_{a_4}=(u_{10},u_{17},u_{18},u_{19},u_{11})$ and
 $P_{a_5}=(u_{11},u_{20},u_{21},u_{22},u_{12})$, respectively
 to obtain a subgraph $G_1$ of $\anC$. 
\item[-]
Next attach to this graph  $G_1$ three new paths 
$Q_{u_5}=(u_5,u_{24})$, 
$Q_{u_{18}}=(u_{18},u_{25},u_{26},u_{27})$ and 
$Q_{u_{22}}=(u_{22},u_{28})$
to obtain  
the interior of  $\anC$ in Figure~\ref{fig:example_chemical_graph}.
\item[-]
Finally  attach to the interior   28 trees selected from the set $\mathcal{F}$ 
and assign chemical elements and bond-multiplicities in the interior
to  obtain a chemical graph $\C$  in Figure~\ref{fig:example_chemical_graph}. 
In Figure~\ref{fig:example_fringe-tree},  
  $\psi_1\in \mathcal{F}$ is selected for $\Co[u_i]$, $i\in\{6,7,11\}$.
    Similarly 
  $\psi_2$  for  $\Co[u_9]$,
  $\psi_4$   for $\Co[u_1]$, 
  $\psi_6$   for $\Co[u_i]$,   
  $i\in\{3,4,5,10,19,22,25,26\}$,
  $\psi_8$   for $\Co[u_8]$, 
  $\psi_{11}$    for $\Co[u_i]$, $i\in\{2,13,16,17,20\}$,
  $\psi_{15}$   for $\Co[u_{12}]$,
   $\psi_{19}$    for $\Co[u_{15}]$,
   $\psi_{23}$    for $\Co[u_{21}]$,
   $\psi_{24}$    for $\Co[u_{24}]$,
   $\psi_{25}$    for $\Co[u_{27}]$, 
   $\psi_{26}$   for $\Co[u_{23}]$, 
   $\psi_{27}$  for $\Co[u_{14}]$     
   and 
    $\psi_{30}$  for $\Co[u_{28}]$. 
\end{enumerate} 

%
%

See Appendix~\ref{sec:specification} for a full description of topological specification.

 \section{Prediction Functions}\label{sec:prediction_functions}  

Let $\mathcal{C}$ be a data set   of chemical graphs $\C$ with
an observed value $a(\C)\in \R$. 
Let $D$ be a set of descriptors with $K=|D|$ and 
 $f$ be a feature function that maps a chemical graph $\C$
to a vector $ f(\C)\in \RK$,
where $x(d)$ denotes the value of descriptor $d\in D$.
For a notational simplicity,
  we denote  $a_i=a(\C_i)$ and  $\x_i= f(\C_i)$ 
for an indexed graph $\C_i\in \mathcal{C}$. 

\subsection{Evaluation}\label{sec:evaluation}  

For  a prediction function $\eta: \RK\to \R$, 
define an error function 
\[ \mathrm{Err}(\eta;\mathcal{C})  \triangleq 
\sum_{\C_i\in \mathcal{C}}(a_i - \eta(f(\C_i)))^2
=\sum_{\C_i\in \mathcal{C}}(a_i - \eta(\x_i))^2, \]
and define the {\em coefficient of determination}
 $\mathrm{R}^2(\eta,\mathcal{C})$ 
  to be 
\[ \displaystyle{ \mathrm{R}^2(\eta,\mathcal{C})\triangleq 
  1- \frac{\mathrm{Err}(\eta;\mathcal{C}) } 
  {\sum_{ \C_i\in \mathcal{C}  } (a_i-\widetilde{a})^2}  
  \mbox{   for  }
   \widetilde{a}= \frac{1}{|\mathcal{C}|}\sum_{ \C\in \mathcal{C} }a(\C).  } \] 
   
We evaluate a method of constructing a prediction function
over a set $D$ of descriptors by 5-fold cross-validation as follows.   
A single run $r$ of 5-fold cross-validation executes the following: 
 Partition a data set $\mathcal{C}$  randomly 
 into five subsets $\mathcal{C}^{(k)}$, $k\in[1,5]$ so that 
 the difference between  $|\mathcal{C}^{(i)}|$ and $|\mathcal{C}^{(j)}|$ is at most 1. 
 For each $k\in[1,5]$, let $\Ctrain:=\mathcal{C} \setminus \mathcal{C}^{(k)}$,
  $\Ctest:=\mathcal{C}^{(k)}$ and
 execute the method to construct a prediction function $\eta^{(k)}:\mathrm{R}^{|D|}\to \R$ over  
 a training set  $\Ctrain$
 and compute $g_r^{(k)}:=\mathrm{R}^2(\eta^{(k)},\Ctest)$.
Let $\mathrm{R}^2_{\mathrm{CV}}(\mathcal{C},D,p)$ denote 
 the median  of $\{g_{r_i}^{(k)}\mid k\in[1,5],i\in[1,p]\}$ of  
 $p$ runs $r_1,r_2,\ldots,r_p$ of 5-fold cross-validation.

\subsection{Linear Regressions}\label{sec:linear_regression}  

For  a set $D$ of descriptors, a hyperplane is defined to be 
a pair  $(\w,b)$ of a vector $\w\in \RK$ and a real $b\in \R$.
Given a hyperplane $(\w,b)$,
a prediction function $\eta_{\w,b}:\RK\to \R$ is defined by setting
\[ \eta_{\w,b}(\x) \triangleq \w\cdot \x +b=\sum_{d\in D}\w(d)\x(d) +b. \]


\paragraph{Multidimensional Linear Regression (MLR)}
Given a data set $\mathcal{C}$ and a set $D$ of descriptors, 
{\em multidimensional linear regression} MLR$(\mathcal{C},D)$ returns
a hyperplane $(\w,b)$ with $\w\in \RK$  
 that minimizes $\mathrm{Err}(\eta_{\w,b}; \mathcal{C})$.
However, such a hyperplane $(\w,b)$ may contain
unnecessarily many non-zero reals $\w(d)$.
To avoid this, a minimization with an additional penalty term $\tau$ 
to the error function has been proposed.
Among them,     
 a Lasso function~\cite{Lasso96} is defined to be 
\[\frac{1}{2|\mathcal{C}|}\mathrm{Err}(\eta_{\w,b}; \mathcal{C})
+   \lambda \tau, ~~~ \tau=\sum_{d\in D}  |w(d)| +  |b |, \]
 where $\lambda \in \R_+$ is a given nonnegative number.


\paragraph{Adjustive Linear Regression (ALR)}
We review a recent learning method, called
{\em adjustive linear regression}, that is effectively
equivalent to an ANN  with no hidden layers by a linear regression
such that each input node may have a non-linear activation function
(see  \cite{ALR_ZHNA22} for the details of the idea). 
 Let $\mathcal{C}=\{\C_1,\C_2,\ldots,\C_m\}$,  $A=\{a_i=f(\C_i)\mid i\in[1,m]\}$ 
 and $X=\{x_i=f(\C_i)\in \RK\mid i\in[1,m]\}$. 
Let $D^+$ (resp., $D^-$) denote the set of descriptor $d\in D$
such that  the correlation coefficient $\sigma(X[d],A)$ 
between $X[d]=\{x_i(d)\mid i\in[1,m]\}$ and  $A$ 
is nonnegative (resp., negative).
We first solve the following linear program
with a constant $\lambda\geq 0$, 
a real variable $b$
and nonnegative real variables
 $c_q(0), q\in [0,2]$, $w_q(d), q\in[0,2], d\in D$.
 
 \smallskip
\noindent 
{Linear Program}
\begin{equation} \label{eq:minimization}
 \begin{array}{l }
\mbox{Minimize: ~}   
 \displaystyle{  \frac{1}{2m}\sum_{i\in[1,m]}
 \Bigl|c_0(0) a_i+c_1(0)  a_i^2+c_2(0)(1-(a_i\!-\! 1)^2)  } \\
    \displaystyle{ ~~~~~~~~~~~~~~~~~~~~~~~~~~~~~
         -  \sum_{d\in D^+}[ \w_0(d) \x_i(d)+ \w_1(d) \x_i(d)^2 +\w_2(d)(1-(\x_i(d)\!-\! 1)^2)]  } \\
    \displaystyle{ ~~~~~~~~~~~~~~~~~~~~~~~~~~~~~
      +  \sum_{d\in D^-}[ \w_0(d) \x_i(d)+ \w_1(d) \x_i(d)^2 + \w_2(d)(1- (\x_i(d)\!-\! 1)^2)]  - b \Bigr| +\lambda\tau } \\
\mbox{subject to}  \\ ~~~~~~~~~~~~~~~~~~~~~ 
 \displaystyle{   \tau = \sum_{d\in D} \w_0(d)   +  |b|,  ~~~
  c_0(0)+c_1(0)+c_2(0)=1.}
\end{array} 
\end{equation} 
An optimal solution to this minimization can be found by solving
a linear program with $O(m+|D|)$ variables and constraints.
From an optimal solution, 
we next compute the following hyperplane $(\w^*,b^*)$ 
to obtain a linear prediction function $\eta_{\w^*,b^*}$. 
Let
 $c^*_q(0), q\in [0,2]$, $w^*_q(d), q\in[0,2], d\in D$ and $b^*$ denote
 the values of variables 
 $c_q(0),  q\in [0,2]$, $w_q(d), q\in[0,2],  d\in D$ and $b$ 
 in an optimal solution, respectively.
 Let  $D^\dagger$ denote the set of descriptors $d\in D$ with $\w^*_0(d)>0$.
 Then we set \\
~~~ $\w^*(d):=0$ for $d\in D$ with $\w^*_0(d)=0$, \\
~~~ $\w^*(d):=\w^*_0(d)/(\w^*_0(d)+\w^*_1(d)+\w^*_2(d))$ for $ d\in D^+\cap D^\dagger$, \\
~~~ $\w^*(d):= - \w^*_0(d)/(\w^*_0(d)+\w^*_1(d)+\w^*_2(d))$ for $ d\in D^-\cap D^\dagger$ and \\
~~~  $\w^*:=(\w^*_0(1),  \w^*_0(2), \ldots, \w^*_0(|D|))\in \RK$. 

\paragraph{Reduction of Descriptors and Linear Regression (RLR)}
We finally review a learning method recently
 proposed by Zhu~et al.~\cite{Q_ZACIHZNA22}
 to improve the learning performance with the two-layered model.
 Given a set of descriptors $x(1),x(2),\ldots,x(K)$,
the method first adds to the original set of linear descriptors 
a quadratic descriptor  $x(i)x(j)$ (or $x(i)(1-x(j))$ of each two descriptors.
This drastically increases the number of descriptors, which would take
extra running time in learning or cause over-fitting to the data set.
Next the method reduces the set of linear and quadratic descriptors
into a smaller set that delivers a prediction function with a higher performance
(see \cite{Q_ZACIHZNA22} for the details on the reduction procedure).
Finally the method constructs a prediction function by using MLR
on the set of selected descriptors.
We call this method based on reduction and linear regression {\em  RLR} in this paper.


\section{Splitting Data Sets via Hyperplanes}\label{sec:LPs}

This section proposes a method of splitting a given data set into two subsets
with a hyperplane in the feature space so that most of the compounds  $\C$ in the first 
(resp., second) subsets 
 have observed values $a(\C)$ smaller (resp., larger) than a threshold $\theta$.
 
 For a property $\pi$, 
 let $\mathcal{C}=\{\C_1, \C_2, \ldots, \C_n\}$ be a set of chemical graphs.
 Assume that all entries and observed values are normalized,
 where $\min\{a_i\mid \C_i\in \mathcal{C}\}=0$ and $\max\{a_i\mid \C_i\in \mathcal{C}\}=1$
 and   $\min\{x_i(d)\mid \C_i\in \mathcal{C}\}=0$ and $\max\{x_i(d)\mid \C_i\in \mathcal{C}\}=1$
 for each $d\in D$.
 
 For a threshold $\theta$ with $0<\theta<1$, we find
 a hyperplane  $(w,b)$ with $w\in \RK$ and $b\in\R$ 
 that splits the set $\mathcal{C}$ into subsets
 \[\mbox{ $\mathcal{C}^{(1)}:=\{\C_i\in \mathcal{C}\mid wx_i-b \leq 0\}$ and 
   $\mathcal{C}^{(2)}:=\{\C_i\in \mathcal{C}\mid wx_i-b >0\}$  }\]
 so that $\mathcal{C}^{(1)}$ (resp., $\mathcal{C}^{(2)}$)
  contains compounds $\C_i\in \mathcal{C}$ with $a_i\leq \theta$ 
  (resp.,   $a_i> \theta$)  as many as possible.
 Then we treat each of the subsets  $\mathcal{C}^{(j)},j=1,2$
 as a new data set   and construct a prediction function  $\psi_j$ before we obtain
 a prediction function to the original set  $\mathcal{C}$
 by combining functions  $\psi_1$ and $\psi_2$. 
 
\paragraph{A Linear Program Formulation to Find a Hyperplane}
 To find a hyperplane to split a given data set, we formulate a linear program 
 in a similar manner of the idea by   Freed and Glover~\cite{FG81} for separating a classification data. 
Define sets $\mathcal{C}_{\leq \theta}:=\{\C_i\in \mathcal{C}\mid a_i\leq \theta\}$
 and $\mathcal{C}_{> \theta}:=\{\C_i\in \mathcal{C}\mid a_i> \theta\}$,
 and choose compounds $\C_{s}\in \mathcal{C}_{\leq \theta}$ with $a_s=0$ 
 and $\C_{t}\in \mathcal{C}_{> \theta}$ with $a_t=1$,
  so that $\C_s\in \mathcal{C}^{(1)}$, $\C_t\in \mathcal{C}^{(2)}$.

 A linear program is formulated as follows, 
 where a hyperplane $(w,b)$ with $w\in \RK$ and $b\in\R$ is obtained 
 as an optimal solution to this  linear program: \\
 
 \medskip\noindent
 {\bf LP}$(\theta)$\\
 {constants}:
  $a_i\in \R$, $x_i\in \RK,~\forall \C_i\in \mathcal{C}$;
  indices $s$ and $t$ such that $a_s=0$ and  $a_t=1$; $\theta\in \R$; \\
 {variables}:
  $w\in \RK$, $b\in\R$, nonnegative variables 
 $\delta_i \geq 0, \forall \C_i\in \mathcal{C}$; \\
 {constraints}:
 \[ w x_s -b\leq 0, \]
 \[w x_t -b\geq 0, \]
 \[ \delta_i\geq  w x_i -b +(a_i-\theta)^2, ~~~ \forall \C_i\in \mathcal{C}_{\leq \theta}, \]
 \[ \delta_i \geq -(w x_i -b)+(a_i-\theta)^2, ~~~ \forall \C_i\in \mathcal{C}_{> \theta}, \]
  {objective function}:   
 \[ \mbox{minimize~~} \sum_{\C_i\in \mathcal{C}} \delta_i.  \]
 
 \medskip
 The above linear program consists of $O(|D|+|\mathcal{C}|)$ variables and 
 $O(|\mathcal{C}|)$ constraints. 
We solve the linear program to obtain  an optimal solution $(w,b)$
and compute $\mathcal{C}^{(1)}=\{\C_i\in \mathcal{C}\mid wx_i-b \leq 0\}$ and 
   $\mathcal{C}^{(2)}=\{\C_i\in \mathcal{C}\mid wx_i-b >0\}$.
Denote
\[ a_{\min}^{(j)}:=\min\{a_i\mid \C_i\in \mathcal{C}^{(j)}\} \mbox{ and } 
  a_{\max}^{(j)}:=\max\{a_i\mid \C_i\in \mathcal{C}^{(j)}\}, j=1,2, \]
  where  $a_{\min}^{(1)}=0$ and $a_{\max}^{(2)}=1$.
When $\mathcal{C}^{(1)}=\mathcal{C}_{\leq \theta}$ and
      $\mathcal{C}^{(2)}=\mathcal{C}_{> \theta}$ hold, 
 the ranges $[a_{\min}^{(1)}=0, a_{\max}^{(1)}]$ and $[a_{\min}^{(2)}, a_{\max}^{(2)}=1]$
 have no overlap (i.e., $a_{\max}^{(1)}\leq \theta < a_{\min}^{(2)}$). 
 Otherwise  $a_{\min}^{(2)} \leq  a_{\max}^{(1)}$ holds, where
 even for this case, the two subsets $\mathcal{C}^{(1)}$ and $\mathcal{C}^{(2)}$
 are well-separated if $a_{\min}^{(2)}$ and $a_{\max}^{(1)}$ are very close.
 We select a threshold $\theta$ from a set of candidates so that 
$a_{\max}^{(1)}- a_{\min}^{(2)}$ is minimized subject to the condition that
 each of $|\mathcal{C}^{(1)}|$ and $|\mathcal{C}^{(2)}|$ becomes
nearly half of the original size $|\mathcal{C}|$.

\paragraph{Implementation in the First Phase of the Framework}
In the first phase of the framework, we construct a  prediction function $\psi$
to a data set $\mathcal{C}$ for a property $\pi$ 
and a descriptor set $D$ of a feature function $f:\mathcal{C}\to \RK$ as follows.
For a selected threshold $\theta$, we find a hyperplane $(w,b)$, $w\in \RK$, $b\in\R$
as an optimal solution to the above linear program {\bf LP}$(\theta)$
based on which we split $\mathcal{C}$ into subsets 
$\mathcal{C}^{(j)}, j=1,2$. 
For each $j=1,2$, 
choose a set $\widetilde{D}_j$ of descriptors and 
 construct a prediction function $\psi_j: \R^{|\widetilde{D}_j|}\to \R$ 
  for the data set $\mathcal{C}^{(j)}$ with the descriptor set $\widetilde{D}_j$.   
In our computational experiments, we use LLR, ANN, ALR and RLR to construct
prediction functions to $\mathcal{C}^{(j)}, j=1,2$ and choose 
as $\psi_j$ one of them with the best learning performance
(where $\widetilde{D}_j$ is a subset of the linear descriptor set $D$ 
when we use LLR, ANN or ALR;
and  $\widetilde{D}_j$ consists of some linear and quadratic descriptors of $D$ 
 when RLR is used to construct $\psi_j$).
 Given a feature vector  $x\in \RK$, 
 use the prediction function $\psi_1$ if $wx-b \leq 0$; and
  use  the prediction function $\psi_2$ otherwise.
 Thus  the prediction function $\psi$ is given by 
\[
     \psi(x):= \left\{\begin{array}{ll}
         \psi_1(x) & \mbox{if $wx-b\leq 0$,} \\
         \psi_2(x)    &  \mbox{otherwise,}   \end{array}
     \right.
\]
where the hyperplane $(w,b)$ is a part of the prediction function $\psi$.

\paragraph{Implementation in the Second Phase of the Framework}
In the second phase of the framework, we need an MILP formulation that
simulates the computing process of a prediction function $\psi$.
Such a formulation for a prediction function constructed with LLR, ANN, ALR or RLR
has been known~\cite{ZAHZNA21,ALR_ZHNA22,Q_ZACIHZNA22}.
For the data set  $\mathcal{C}$ for property $\pi$ in the first phase, 
let  $(w,b)$ denote the hyperplane that splits $\mathcal{C}$ into 
subsets $\mathcal{C}^{(j)}, j=1,2$ 
and $\psi_j$ be a prediction function 
constructed for $\mathcal{C}^{(j)}$. 
Assume that, for the feature function $f$, a topological specification $\sigma$ and
 each prediction function $\psi_j,j=1,2$, we have  
an MILP formulation  $\mathcal{M}_{f,\eta_j,\sigma}$ for inferring a chemical graph
that satisfies $\sigma$  in the second phase.

Let $\underline{y}^*$ and $\overline{y}^*$ be lower and upper limits for a target value to property $\pi$.
Recall that the observed value $a(\C)$ of a chemical compound $\C\in \mathcal{C}$ is
normalized to a value $\nu(a(\C))$ between 0 and 1.
Let $\nu(\underline{y}^*)$ and $\nu(\overline{y}^*)$ denote
the normalized values of $\underline{y}^*$ and $\overline{y}^*$, respectively, 
where we  assume that either  
$\nu(\overline{y}^*)\leq \max\{a_{\max}^{(1)},\theta\}$ or 
$\min\{a_{\min}^{(2)},\theta\} \leq \nu(\underline{y}^*)$ 
(otherwise we consider two target instances
 $[\nu(\underline{y}^*),\max\{a_{\max}^{(1)},\theta\}]$
 and $[\min\{a_{\min}^{(2)},\theta\},\nu(\overline{y}^*)]$).
In the former (resp., the latter), 
we solve  the MILP $\mathcal{M}_{f,\eta_1,\sigma}$ plus an additional constraint of  $wx_i-b \leq 0$ 
(resp., MILP $\mathcal{M}_{f,\eta_2,\sigma}$ plus an additional constraint of  $wx_i-b \geq 0$) 
to infer a desired chemical graph $\C^\dagger$.


\section{Results}\label{sec:experiment}

With our new method of splitting a data set and
formulating an MILP to treat quadratic descriptors in the two-layered model,
we implemented the framework
for inferring chemical graphs  and
conducted experiments  to evaluate the computational efficiency. 
We executed the experiments on a PC with 
 Processor:  Core i7-9700 (3.0GHz; 4.7 GHz at the maximum) and 
Memory: 16 GB RAM DDR4. 
To construct a prediction function by MLR (multidimensional linear regression), 
or ANN (artificial neural network), 
we used {\tt scikit-learn} version 1.0.2  with Python 3.8.12, 
MLPRegressor for MLR,
 and ReLU activation function for ANN. 

\subsection{Results on the First Phase of the Framework}

\noindent
 {\bf Chemical Properties}
We implemented the first phase for the following  
22 chemical properties of monomers: \\ 
~ autoignition temperature ({\sc At}), 
   biological half life ({\sc BHL}), 
   critical pressure  ({\sc Cp}),  \\
~  critical temperature ({\sc Ct}),
  dissociation constants ({\sc Dc}),     
  flammable limits lower ({\sc FlmL}), \\
~ flammable limits upper ({\sc FlmU}), 
  flash point in closed cup ({\sc Fp}),  
  melting point  ({\sc Mp}),   \\
~ refractive index of trees ({\sc RfIdT}), 
     odor threshold lower  ({\sc OdrL}),  
    odor threshold upper  ({\sc OdrU})  \\
~ and  vapor pressure  ({\sc Vp}) \\
~  provided  by HSDB from PubChem~\cite{pubchem}; \\  
~ solubility ({\sc Sl}) by ESOL~\cite{moleculenet}; \\ 
~  autoignition temperature for organic compounds ({\sc AtO})  
   by A. Dashti et al.~\cite{DJAM20}; \\
~ flammable limits upper for organic compounds ({\sc FlmUO})  
 by S.~Yuan et al.~\cite{YJQKM19}; \\ 
~  flammable limits lower for gas ({\sc FlmLG}) and \\
~ flammable limits upper for gas ({\sc FlmUG}) 
  by S.~Kondo et al.~\cite{KUTTT01};
  and \\  
~ energy of highest occupied molecular orbital ({\sc Homo}),  \\
~  energy of lowest unoccupied molecular orbital ({\sc Lumo}),   \\
~  the energy difference between  {\sc Homo} and {\sc Lumo} ({\sc Gap}) and    \\
~ electric dipole moment ({\sc mu})   provided by  MoleculeNet~\cite{QM9},  
where all these from {\sc Homo} \\
~ to {\sc mu}   are based on a common data set QM9.

 The data set  QM9 contains more than 130,000 compounds.
 In our experiment, we use a set of  1,000 compounds randomly selected from the data set. 
We do not exclude any polymer from the original data set as outliers for these properties.

\begin{table}[h!]\caption{Results of setting data sets.} 
  \begin{center}
    \begin{tabular}{@{} c c r c  c  r r r   @{}}\hline
      $\pi$ & $\Lambda$  &  $|\mathcal{C}_{\pi}|$  & 
       $ \underline{n},~\overline{n} $ &   $\underline{a},~\overline{a}$ &
   $|\Gamma|$   &  $|\mathcal{F}|$ &   $K_1$ \\ \hline 
      {\sc At} &  $\lambda_{1}$  & 400  &  4,\,85  &  64.0,\,715.0    & 23   & 160  & 216 \\
      {\sc At} &   $\lambda_{3}$  & 448   &  4,\,85  &   64.0,\,715.0   &  28  & 181  & 254 \\
      {\sc AtO} &   $\lambda_{5}$  & 443  &  2,\,32  &  170.0,\,680.0    & 16   & 205  & 262 \\
      {\sc BHL} &   $\lambda_{1}$  & 300  &  5,\,36  &   -1.522,\,2.865   &   20 &  70 & 117 \\
      {\sc BHL} & $\lambda_{3}$  & 514  &  5,\,36  &  -1.522,\,2.865   & 26  & 101  & 164    \\
      {\sc Cp} &   $\lambda_{1}$  & 125   &  4,\,63  &  $4.7\!\times\! 10^{-6}$,\,5.52    &  8  & 75  & 107 \\
      {\sc Cp} & $\lambda_{4}$  & 131  &  4,\,63  &  $4.7\!\times\! 10^{-6}$,\,5.52  & 8   & 79  & 115      \\  
       {\sc Ct} & $\lambda_{1}$  &  125  & 4,\,63   &  56.1,\,3607.5    &  8  &  76 &  108 \\
       {\sc Ct} & $\lambda_{4}$  &  132 &  4,\,63 &   56.1,\,3607.5  &  8  & 81   & 117 \\
      {\sc Dc} & $\lambda_{1}$  & 141  &  5,\,44  &  0.5,\,17.11   & 20   & 62  & 109     \\ 
      {\sc Dc} & $\lambda_{3}$  & 161  &  5,\,44  &  0.5,\,17.11   & 25   & 69  & 128     \\   
      {\sc FlmL}  & $\lambda_{6}$   & 254   & 4,\,67   & -0.585,\,0.875  &  19   & 126   & 177 \\  
      {\sc FlmLG}  & $\lambda_{8}$   & 233   & 1,\,13   & -0.221,\,1.158  &  10   & 152   & 199 \\  
      {\sc FlmU}  & $\lambda_{6}$   & 219   & 4,\,67 & 0.107,\,1.681   &  19     & 119   & 170  \\  
      {\sc FlmUO}  & $\lambda_{7}$   & 78   & 2,\,10   & 0.732,\,1.903  &  7   & 61   & 99 \\  
      {\sc FlmUG}  & $\lambda_{7}$   & 233  & 1,\,13   & 0.462,\,2.0  &  10   & 152  & 199 \\  
      {\sc Fp} & $\lambda_{1}$  & 368  &  4,\,67  &  -82.99,\,300.0   & 20   & 131  & 181      \\ 
      {\sc Fp} & $\lambda_{3}$  & 424  &  4,\,67  &  -82.99,\,300.0   & 25   & 161  & 228       \\    
      {\sc Mp} & $\lambda_{1}$ & 467 &4,\,122 &   -185.33,\,300.0  & 23 &142   & 195 \\
      {\sc Mp} & $\lambda_{3}$ & 577 &4,\,122 &   -185.33,\,300.0  & 32 &176   & 253 \\ 
      {\sc OdrL} & $\lambda_{1}$  & 64  & 4,\,13  & 0.0002,\,725.0    &  13  &  49 & 88      \\ 
      {\sc OdrL} & $\lambda_{3}$  & 83 &   4,\,22  & 0.0002,\,725.0   &  16   & 60  &   107    \\  
      {\sc OdrU} & $\lambda_{1}$  & 64  & 4,\,13   & 0.024,\,6000.0     & 13    & 49   & 88      \\ 
      {\sc OdrU} & $\lambda_{3}$  & 83   & 4,\,22   & 0.015,\,6000.0     & 16   & 60  & 107      \\  
      {\sc RfIdT} &   $\lambda_{1}$  & 166  & 4,\,26 &  1.3326,\,1.613 &  14 &  98 &  139 \\
      {\sc Sl} & $\lambda_{1}$  & 673  &  4,\,55  &  -9.332,\,1.11   & 27   & 154  & 216     \\ 
      {\sc Sl} & $\lambda_{3}$  & 915 &  4,\,55  &  -11.6,\,1.11  & 42   & 207  & 299      \\  
      {\sc Vp} & $\lambda_{1}$  &  392 & 4,\,55    &  -8.0,\,3.416  &  22   &133   &  185     \\  
      {\sc Vp} & $\lambda_{3}$  & 482 & 4,\,55    &  -8.0,\,3.416  &  30   & 165  &  236     \\  
      {\sc Homo} &  $\lambda_{2}$  & 977 &  6,\,9   &-0.3335,\, -0.1583  & 59  & 190  & 296 \\
      {\sc Lumo} & $\lambda_{2}$ & 977  &  6,\,9  &  -0.1144,\,0.1026   & 59   & 190 & 296   \\
      {\sc Gap} & $\lambda_{2}$  &  977  &  6,\,9   & 0.1324,\,0.4117  & 59 &  190  & 296 \\  
       {\sc mu} & $\lambda_{2}$ & 977 & 6,\,9   &   0.04,\,6.8966   & 59  &190   & 296 \\   
      \hline
  \end{tabular}\end{center}\label{table:phase1a2}
\end{table}

 \medskip \noindent
 {\bf Setting Data Sets}
For each property $\pi$,  
 we first select a set $\Lambda$ of chemical elements 
 and then collect  a  data set  $\mathcal{C}_{\pi}$ on chemical graphs
 over the set $\Lambda$ of chemical elements.  
 To construct the data set $\mathcal{C}_{\pi}$,
  we eliminated  chemical compounds that do not satisfy 
  one of the following: the graph is connected,
  the number of carbon atoms is at least four,
  and   the number of non-hydrogen neighbors of each atom is
  at most 4.   

We set a branch-parameter ${\rho}$ to be 2, 
introduce linear   descriptors defined by the two-layered graph 
in the chemical model without suppressing hydrogen 
 and use the sets $D_\pi^{(1)}$ and $D_\pi^{(2)}$ 
 of linear and quadratic descriptors  
   (see Appendix~\ref{sec:descriptor} for the details).

 For properties $\pi\in\{${\sc BHL, FlmL, FlmLG, FlmU, FlmUO, FlmUG, OdrL, OdrU, Vp}$\}$, 
 we take the logarithmic measurement of  observed values $a(\C), \C\in\mathcal{D}_{\pi}$, where 
  $\log (a(\C)+c)$ ($c=10^{-8}$ if $\pi=${\sc Vp} and $c=0$ otherwise)    
  is used as the observed value in our experiments. 

 We normalize the range of each linear descriptor and
 the range  of observed values   $a(\Co), \Co\in \mathcal{C}_\pi$.
 
 We conducted an experiment of  comparing the following five methods of constructing a prediction function.
 \begin{enumerate}
 \item[(i)] {\bf LLR}: use Lasso linear regression on the  set    $D_\pi^{(1)}$ of linear descriptors
 (see \cite{ZAHZNA21} for the detail of the implementation); 
 \item[(ii)]  {\bf ANN}:  use ANN on the  set  $D_\pi^{(1)}$ of linear descriptors
 (see \cite{ZAHZNA21} for the detail of the implementation); 
 \item[(iii)]  {\bf ALR}:  use adjustive linear regression on the  set   $D_\pi^{(1)}$ of linear descriptors
 (see \cite{ALR_ZHNA22} for the detail of the implementation);  
 \item[(iv)]  {\bf RLR}:  the learning method proposed by Zhu~et al.~\cite{Q_ZACIHZNA22}
 that chooses a set of descriptors from
 the set $D_\pi^{(1)}\cup D_\pi^{(2)}$ of linear and quadratic descriptors and
 then constructs a prediction function by applying MLR (multidimensional linear regression)
  to the resulting set of descriptors; and
 \item[(v)]  {\bf HPS}:  
   our method of computing a hyperplane to split a given data set
   into two subsets 
   and constructing a prediction function to each subset independently. 
 By conducting a preliminary experiment,
 we predetermine a threshold $\theta$ and a hyperplane $(w,b), w\in \R^{|D_\pi^{(1)}|}$ and $b\in \R$
in the feature space of linear descriptors.
In a cross-validation, we split a training data set $\widehat{\mathcal{C}}_\pi \subseteq \mathcal{C}_\pi$ 
into two subsets $\widehat{\mathcal{C}}_\pi^{(j)}, j=1,2$ and construct 
 a prediction function $\psi_j$ to each subset $\widehat{\mathcal{C}}_\pi^{(j)}$
 by applying one of the above four methods (i)-(iv).  
\end{enumerate}

 Table~\ref{table:phase1a2} shows the size and range of data sets   that 
 we prepared for each chemical property to construct a prediction function,
 where  we denote the following:  
\begin{enumerate}[nosep,  leftmargin=*]
\item[-] $\pi$: the name of a chemical property used in the experiment.
\item[-] 
  $\Lambda$: a set of selected elements used in the data set $\mathcal{C}_{\pi}$; 
  $\Lambda$ is one of the following eight sets:  \\
$\Lambda_1=\{\ttH,\ttC,\ttO, \ttN \}$;
$\Lambda _2=\{\ttH,\ttC,\ttO, \ttN, \ttF \}$; 
 $\Lambda_3=\{\ttH,\ttC,\ttO, \ttN, \ttCl, \ttS_{(2)}, \ttS_{(6)} \}$; 
 $\Lambda_4=\{\ttH,\ttC,\ttO, \ttN, \ttCl, \ttPb \}$;
 $\Lambda_5=\{\ttH,\ttC,\ttO, \ttN,\ttS_{(2)},\ttS_{(4)},\ttS_{(6)},\ttCl ,\ttBr, \ttF \}$; 
 $\Lambda_6=\{\ttH,\ttC,\ttO, \ttN, \ttCl ,\ttP_{(2)},\ttP_{(5)} \}$;
 $\Lambda_7=\{\ttH,\ttC,\ttO, \ttN, \ttCl ,\ttBr \}$;  
 $\Lambda_8=\{\ttH,\ttC,\ttO, \ttN, \ttCl ,\ttBr, \ttF \}$,
 where ${\tt a}_{(i)}$ for a chemical element ${\tt a}$ and an integer $i\geq 1$ 
 means that  a chemical element ${\tt a}$ with valence $i$. 
 
\item[-] 
 $|\mathcal{C}_{\pi}|$:  the size of data set $\mathcal{C}_{\pi}$ over the element set $\Lambda$
  for the property $\pi$.
   
\item[-]   $ \underline{n},~\overline{n} $:  
  the minimum and maximum  values of the number 
  $n(\Co)$ of non-hydrogen atoms in 
  the   compounds $\Co$ in $\mathcal{C}_{\pi}$.
\item[-] $ \underline{a},~\overline{a} $:  the minimum and maximum values
of $a(\Co)$ (or the logarithm of the original observed values)
for $\pi$ over   the   compounds $\Co$ in  $\mathcal{C}_{\pi}$
before we normalize them between 0 and 1.
\item[-]    $|\Gamma|$: 
the number of different edge-configurations
of interior-edges over the compounds in~$\mathcal{C}_{\pi}$. 
\item[-]  $|\mathcal{F}|$: the number of non-isomorphic chemical rooted trees
 in the set of all 2-fringe-trees in  the   compounds in $\mathcal{C}_{\pi}$.
 
\item[-]  $K_1$: the size  $|D_\pi^{(1)}|$ of a set $D_\pi^{(1)}$ of linear descriptors,
where $|D_\pi^{(2)}|=  (3K_1^2+K_1)/2$ holds.  
\end{enumerate}

\medskip \noindent
{\bf Constructing Prediction Functions}
For each chemical property $\pi$, we construct a prediction function
by one of the four methods (i)-(iv). 

For methods (i)-(iv), we used the same implementation  
due to Zhu~et al.~\cite{ZAHZNA21,ALR_ZHNA22,Q_ZACIHZNA22}
and omit the details.

 Tables~\ref{table:phase1} shows  
 the results on constructing prediction functions,
 where  we denote the following:     
\begin{enumerate}[nosep,  leftmargin=*]  
\item[-] $\pi, \Lambda$: an instance with a chemical property $\pi$
and a set $\Lambda$ of chemical elements selected from the data set $\mathcal{C}_{\pi}$. 

\item[-] LLR:
the median of test $\mathrm{R}^2$  
  in ten 5-fold cross-validations for prediction functions constructed by the method (i). 
  
\item[-] ANN:
the median of test $\mathrm{R}^2$  
  in ten 5-fold cross-validations for prediction functions constructed by the method (ii).  
  
\item[-] ALR:
the median of test $\mathrm{R}^2$  
  in ten 5-fold cross-validations for prediction functions constructed by the method (iii).  
  
\item[-] RLR:
the median of test $\mathrm{R}^2$  
  in ten 5-fold cross-validations for prediction functions constructed by the method (iv).  
  
\item[-]   HPS:   
the median of test $\mathrm{R}^2$  
  in ten 5-fold cross-validations for the prediction function obtained
  by   HPS, where a prediction function is obtained by combining
  the prediction functions constructed for the first and second subsets.
  
\item[-] the score of LLR, ANN, ALR, RLR or HPS  marked with ``*'' indicates
the best performance among the five methods for the property $\pi$; 
  
\item[-]  $\theta$: the threshold used in HPS to split the data set $\mathcal{C}_{\pi}$.

\item[-]  $a^{(1)}_{\max},a^{(2)}_{\min}$:
the maximum observed value $a(\C)$ of a compound $\C\in \mathcal{C}^{(1)}$ and
the minimum observed value $a(\C)$ of a compound $\C\in \mathcal{C}^{(2)}$
 in HPS:  

\item[-] $|\mathcal{C}^{(1)}|,|\mathcal{C}^{(2)}|$:  
the sizes of the first subset $\mathcal{C}^{(1)}$ and the second subset $\mathcal{C}^{(2)}$
 in HPS.

\item[-]   $\mathcal{C}^{(1)}$-R$^2$:
the name of the method (one of (i)-(iv))
 used to construct a prediction function to the first subset  in HPS and 
the median of test $\mathrm{R}^2$  
  in ten 5-fold cross-validations for the prediction function over the first subset.  

\item[-]    $\mathcal{C}^{(2)}$-R$^2$:  
the name of the method  (one of (i)-(iv))
 used to construct a prediction function to the second subset  in HPS and 
the median of test $\mathrm{R}^2$  
  in ten 5-fold cross-validations for the prediction function over the second subset.  
\end{enumerate}

\bigskip 
The running time of  constructing two prediction functions for subsets $\mathcal{C}^{(1)}$ and $\mathcal{C}^{(2)}$
in HPS was at most around 13 seconds.
To execute RLR,  selecting a subset of linear and quadratic descriptors 
is the most time consuming part, which took from 1162 to 44356 seconds.


\begin{table}[h!]\caption{Results of constructing prediction functions for monomers.} 
  \begin{center}
\scalebox{0.83}{
    \begin{tabular}{@{} l  c c c r r c c c c c c c c  @{}}\toprule
      ~$\pi, \Lambda$  & LLR   & ANN & ALR  & RLR & ~HPS   &  $\theta$~   &  $a^{(1)}_{\max},a^{(2)}_{\min}$
        & \,{\small $|\mathcal{C}^{(1)}|,|\mathcal{C}^{(2)}|$}  &    $\mathcal{C}^{(1)}$-R$^2$
        &   $\mathcal{C}^{(2)}$-R$^2$     
         \\ \midrule  
      {\sc At},   $\lambda_{1}$ & 0.363 & 0.465  & 0.449  & 0.505 &  *0.808   & 0.40 &0.505, 0.293 & 147, 253~  & RLR 0.220 & RLR  0.521\\
     {\sc At},    $\lambda_{3}$ & 0.391 &0.476 &0.442& 0.501 & *0.765  & 0.35 & 0.555, 0.178 & 121, 327~ & ANN 0.126 & RLR  0.532  \\
       {\sc AtO},     $\lambda_{5}$ & 0.710&0.715&0.716& 0.829 & *0.894  & 0.45& 0.447, 0.451 &223, 220~ & ANN 0.747& RLR  0.610  \\
     {\sc BHL},    $\lambda_{1}$   &  0.580 & 0.648 & 0.679& 0.759  &  *0.818  & 0.45 & 0.598, 0.259 &124, 176~  &  RLR  0.650  & RLR  0.560   \\
     {\sc BHL},  $\lambda_{3}$ &  0.688 & 0.751 & 0.706 & *0.834 &  0.831 & 0.50 & 0.753, 0.183 & 240, 274~ & RLR  0.620  & RLR  0.569  \\
       {\sc Cp},   $\lambda_{1}$   &  0.429 & 0.592& 0.805& 0.677& *0.887  & 0.55 & 0.549, 0.562  &  59,\,~66~ & RLR   0.759  & RLR   0.632   \\
        {\sc Cp},   $\lambda_{4}$   & 0.559 & 0.768 &0.546& 0.841  & *0.850  &0.60 & 0.598, 0.601  & 68, \,63~ & RLR  0.661 & RLR   0.564  \\
      {\sc Ct},    $\lambda_{1}$   &  0.069 &  0.290  & 0.903  & *0.937 &  0.863  &   0.15 & 0.150, 0.151  &60, \,65~ & RLR   0.713  & LLR   0.740  \\
      {\sc Ct},    $\lambda_{4}$   &  0.037 & 0.236 & *0.941 & 0.860  &  0.920 &   0.15 & 0.150, 0.150  &61, \,71~ & RLR   0.791  & RLR   0.883   \\
      {\sc Dc},    $\lambda_{1}$   & 0.559 & 0.662 &0.529& 0.908 & *0.939  & 0.45  & 0.440, 0.456  & 68, \,73~ & RLR 0.787 & RLR 0.810     \\
      {\sc Dc},    $\lambda_{3}$   &  0.574 & 0.628&0.534& 0.829 & *0.919  & 0.40 & 0.386, 0.400  & 78, \,83~ & RLR  0.551  &RLR   0.760    \\   
     {\sc FlmL},   $\lambda_{6}$   & 0.412 & 0.524&0.707 &0.604 & *0.833 & 0.50& 0.499, 0.502 & 106, 113~ & RLR  0.920 & RLR  0.363   \\
    {\sc FlmLG},   $\lambda_{8}$   & 0.850 & 0.786&0.824 & 0.925  & *0.935 & 0.25 & 0.243, 0.254 & 38, \,40~ & RLR  0.838 & RLR  0.833   \\
     {\sc FlmU},  $\lambda_{6}$   & 0.146  &  0.295 & 0.311 & 0.538 & *0.703  & 0.45 & 0.445, 0.451 & 167,\,\,\,66~\,\, &  RLR 0.477 & RLR 0.344 \\
     {\sc FlmUO},   $\lambda_{7}$ & 0.442 &0.573 &0.221& 0.642 & *0.851 & 0.45& 0.429, 0.455 & 107, 147~ & RLR  0.945  &RLR   0.559  \\
     {\sc FlmUG},   $\lambda_{7}$   & 0.556 & 0.649&0.443& 0.655 & *0.837 & 0.35& 0.346, 0.363 &126, 107~ & RLR 0.679 & RLR  0.292   \\
      {\sc Fp},   $\lambda_{1}$   &  0.607& 0.742& 0.653& 0.899 & *0.920 &0.40 & 0.399, 0.402 &178, 190~ & RLR  0.684 & RLR   0.880   \\
     {\sc Fp},   $\lambda_{3}$   &  0.593& 0.634& 0.626& 0.846 & *0.880 & 0.45 & 0.449, 0.452 &261, 163~ & RLR  0.697  &RLR   0.645    \\ 
    {\sc  Mp},    $\lambda_{1}$   &  0.815& 0.840& 0.850&0.873  & *0.923 & 0.50 & 0.493, 0.504  &272, 195~ & RLR  0.607  &RLR   0.778     \\
    {\sc  Mp},     $\lambda_{3}$ &  0.786&0.843& 0.803 & 0.898 & *0.920 & 0.55 & 0.547, 0.542  &329, 248~ & ANN  0.659 & RLR  0.747     \\
   {\sc OdrL},    $\lambda_{1}$   &  -0.034\, & -0.402\,&-0.098\,& 0.449 & *0.901 & 0.50 & 0.498, 0.503 &29, \,35~ & RLR  0.695 & RLR  0.836  \\
  {\sc OdrL},   $\lambda_{3}$   &  -0.055\, & -0.172\, & 0.041& 0.532& *0.827 & 0.50& 0.498, 0.503 &38, \,45~ & RLR  0.543 & RLR 0.657   \\
   {\sc OdrU},    $\lambda_{1}$    & 0.008 &0.170& 0.214 & 0.641& *0.931 & 0.55 &0.541, 0.550 &32, \,32~ & RLR 0.755 &RLR 0.800  \\
   {\sc OdrU},   $\lambda_{3}$  & 0.164 & 0.266 & 0.392 & 0.654 & *0.955 & 0.50& 0.496, 0.500 &40, \,43~ & RLR  0.936  & RLR 0.749   \\
    {\sc  RfIdT},    $\lambda_{1}$   &  0.679&0.770& 0.677& *0.876 & *0.876 & 0.30 & 0.294, 0.303 &89, \,77~ & RLR   0.873  & RLR   0.594     \\ 
   {\sc  Sl},    $\lambda_{1}$   &  0.771& 0.831& 0.788& 0.894 & *0.923 & 0.45 & 0.449, 0.439 &109, 564~ & RLR   0.781  &RLR   0.847  \\
   {\sc Sl},    $\lambda_{3}$   &  0.807& 0.867& 0.813& 0.897 & *0.916 & 0.50& 0.548, 0.500  &128, 787~ & RLR  0.859  &RLR   0.832      \\
  {\sc Vp},    $\lambda_{1}$   &  0.871 & 0.937 &0.893 &0.969& *0.986 & 0.45 &0.449, 0.453 &155, 237~  & RLR 0.930 & RLR 0.949   \\
  {\sc Vp},    $\lambda_{3}$   & 0.830 &0.922& 0.871 & 0.959& *0.977  & 0.45 &0.449, 0.453 &210, 272~  & RLR 0.845 & RLR  0.935   \\
     {\sc Gap},    $\lambda_{2}$   &  0.784 & 0.763&0.744 & 0.876  & *0.907 & 0.25 & 0.298, 0.162  &131, 846~ & RLR   0.524  &RLR  0.863     \\
      {\sc Homo},    $\lambda_{2}$  & 0.704 & 0.608 &0.657 & 0.804  & *0.847 & 0.65  &0.792, 0.557  &849, 128~ & RLR   0.733 & RLR  0.768    \\
     {\sc Lumo},   $\lambda_{2}$  &  0.841 & 0.843&0.819 & 0.920 & *0.948  & 0.70 & 0.700, 0.701  &700, 277~ & RLR 0.874  & RLR  0.855   \\
      {\sc mu},   $\lambda_{2}$  &0.366  & 0.442 &0.399& 0.645  & *0.708 &0.20 & 0.506, 0.021  &169, 808~ & RLR   0.708  & RLR   0.593    \\
      \bottomrule
  \end{tabular} }
  \end{center}\label{table:phase1}
\end{table} 

In Table~\ref{table:phase1},   HPS attains the best score of the median test R$^2$
among the five methods in most cases.
Especially the improvement over the methods (i)-(iv) is significant for properties 
{\sc At, FlmL, FlmU, FlmUO, FlmUG, OdrL} and {\sc OdrU}. 

From the values $a^{(1)}_{\max},a^{(2)}_{\min}$ in Table~\ref{table:phase1},
we see how the original set $\mathcal{C}_\pi$ is split with a hyperplane
into two subsets $\mathcal{C}^{(j)}, j=1,2$ for each property $\pi$.
For property {\sc At} with $\lambda_{1}$, it holds that
$a^{(1)}_{\max}=0.505>a^{(2)}_{\min}=0.293$, which implies that 
 no hyperplane separates 
$\mathcal{C}_{\leq \theta}=\{\C_i\in \mathcal{C}_\pi \mid a_i\leq \theta\}$
 and $\mathcal{C}_{> \theta}=\{\C_i\in \mathcal{C}_\pi \mid a_i> \theta\}$ for $\theta=0.40$.
 On the other hand, for property {\sc AtO}, it holds that
$a^{(1)}_{\max}=0.447<\theta=0.45<a^{(2)}_{\min}=0.451$, which implies that 
 $\mathcal{C}_\pi$ is split with a hyperplane into 
$\mathcal{C}^{(1)}=\mathcal{C}_{\leq \theta}$
 and $ \mathcal{C}^{(2)}=\mathcal{C}_{> \theta}$.

For some properties   such as {\sc At} with $\lambda_{1}$, the median R$^2$ of HPS
 is considerably larger than the median R$^2$ of
   $\mathcal{C}^{(1)}$-R$^2$ and   $\mathcal{C}^{(2)}$-R$^2$.
For property $\pi=$~{\sc At} with  $\lambda_{1}$,      the median R$^2$ of HPS is 0.808  
whereas that of  $\mathcal{C}^{(1)}$ (resp., $\mathcal{C}^{(2)}$) is   0.220  (resp., 0.521).
This can happen because the range $[a_{\min}^{(1)}=0, a_{\max}^{(1)}=0.505]$ 
(resp., $[a_{\min}^{(2)}=0.293, a_{\max}^{(2)}=1]$) 
of observed values for  $\mathcal{C}^{(1)}$ (resp., $\mathcal{C}^{(2)}$)
is again normalized to $[0,1]$ 
on which a prediction function $\psi_1$ (resp., $\psi_2$)
 is constructed and its learning performance
  $\mathcal{C}^{(1)}$-R$^2$ (resp., $\mathcal{C}^{(2)}$-R$^2$) 
is evaluated.
However, in the evaluation of $\psi$ for the median R$^2$ of HPS,
 the error caused by each of the prediction functions $\psi_j, j=1,2$ 
is measured as a relatively smaller value
 over the original wider  range of observed values to $\mathcal{C}_\pi$.


\subsection{Results on the Second Phase of the Framework}

To execute the second phase, 
we used a set of seven instances
$I_{\mathrm{a}}$, $I_{\mathrm{b}}^i, i\in[1,4]$, $I_{\mathrm{c}}$
 and $I_{\mathrm{d}}$ based on the seed graphs prepared by Zhu et~al.~\cite{ZAHZNA21}. 
The instances  $I_{\mathrm{a}}$ and $I_{\mathrm{c}}$ have restricted seed graphs,
the   instances $I_{\mathrm{b}}^i$ have abstract seed graphs and
instances $I_{\mathrm{c}}$ and $I_{\mathrm{d}}$ have restricted set of fringe-trees.
We here present their seed graphs $\GC$ 
(see Appendix~\ref{sec:specification} for the details of $I_{\mathrm{a}}$
and Appendix~\ref{sec:test_instances} for the details of 
$I_{\mathrm{b}}^i, i\in[1,4]$, $I_{\mathrm{c}}$  and $I_{\mathrm{d}}$).

\begin{figure}[h!] \begin{center}
\includegraphics[width=.85\columnwidth]{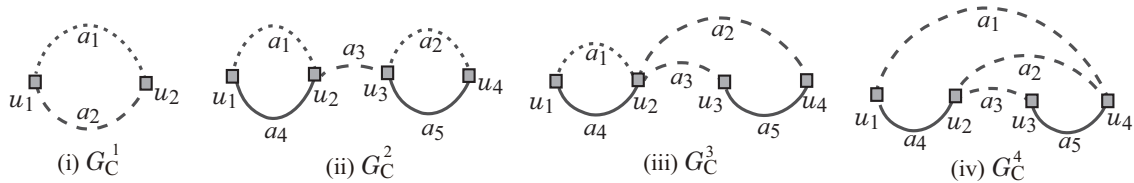}
\end{center} \caption{
(i)  Seed graph   $\GC^1$ for $I_{\mathrm{b}}^1$ and  $I_{\mathrm{d}}$;
(ii) Seed graph $\GC^2$   for $I_{\mathrm{b}}^2$; 
(iii) Seed graph  $\GC^3$   for $I_{\mathrm{b}}^3$; 
(iv)  Seed graph $\GC^4$  for $I_{\mathrm{b}}^4$. }
\label{fig:specification_example_polymer}
\end{figure}

The seed graph  $\GC$ of  $I_{\mathrm{a}}$ is given
 by the graph in Figure~\ref{fig:specification_example_1}(a).
The seed graph $\GC^1$ of  $I_{\mathrm{b}}^1$
(resp., $\GC^i, i=2,3,4$ of $I_{\mathrm{b}}^i,  i=2,3,4$) is illustrated
 in Figure~\ref{fig:specification_example_polymer}.
 
\begin{figure}[!htb]
\begin{center} 
 \includegraphics[width=.69\columnwidth]{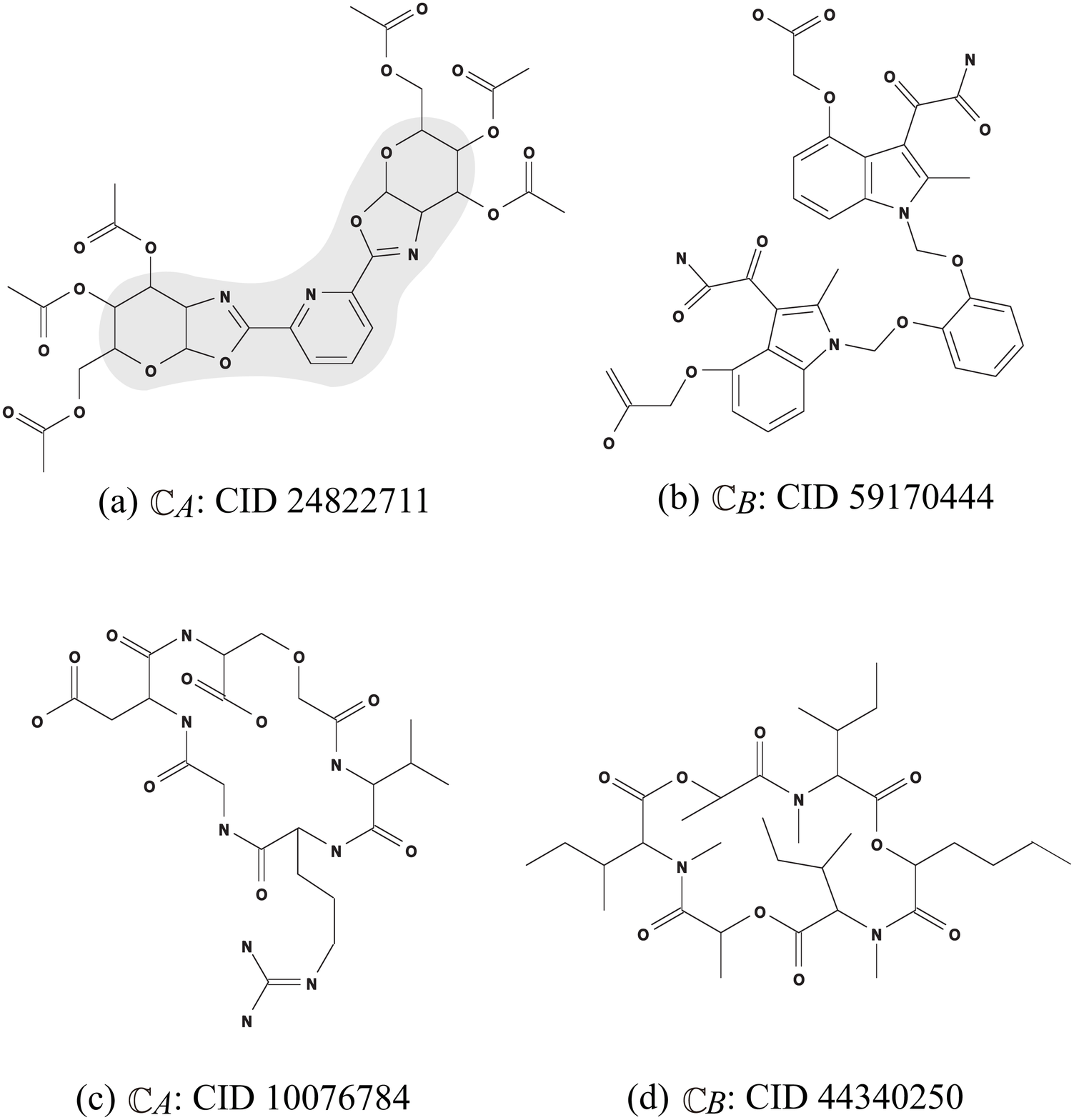}
\end{center}
\caption{An illustration of  chemical compounds 
  for instances  $I_{\rm c}$  and  $I_{\rm d}$: 
(a) $\Co_A$: CID~24822711;
(b)  $\Co_B$: CID~59170444; 
(c) $\Co_A$: CID~10076784;
(d)  $\Co_B$: CID~44340250,
where hydrogens are omitted. 
}
\label{fig:instance_I_c_I_d}  
\end{figure}

Instance  $I_{\mathrm{c}}$ has been introduced 
in order to infer a chemical graph $\Co^\dagger$ such that \\ 
- a core part of $\Co^\dagger$  is equal to that of 
chemical graph $\Co_A$: CID~24822711 in Figure~\ref{fig:instance_I_c_I_d}(a) \\
~ (where the seed graph  $\GC$ of   $I_{\mathrm{c}}$ is indicated 
by the shaded area in Figure~\ref{fig:instance_I_c_I_d}(a)). \\
-  the frequency of each edge-configuration in the non-core of $\Co^\dagger$
is equal to that of chemical graph \\
~   $\Co_B$:  CID~59170444  in  Figure~\ref{fig:instance_I_c_I_d}(b).

Instance  $I_{\mathrm{d}}$ has been introduced 
in order to   infer a chemical  graph $\Co^\dagger$ such that \\
-   $\Co^\dagger$ is monocyclic (where 
the seed graph  of    $I_{\mathrm{d}}$  is given by  $\GC^1$  
 in Figure~\ref{fig:specification_example_polymer}(i)); and  \\
-  the frequency vector of  edge-configurations in  $\Co^\dagger$
is a vector obtained by merging those of \\
~  chemical graphs $\Co_A$: CID~10076784   and $\Co_B$: CID~44340250 
in   Figure~\ref{fig:instance_I_c_I_d}(c) and (d), respectively.   

\medskip \noindent
{\bf Solving an MILP for the Inverse Problem}   
We executed the stage of solving an MILP to infer a chemical graph
 for two properties  $\pi\in \{${\sc At,  FlmL}$\}$.

 For the MILP formulation  $\mathcal{M}_{f,\eta,\sigma}$,
we use the prediction function  $\eta$ for each of 
 {\sc At} with $\Lambda_3$ and {\sc FlmL} with $\Lambda_6$ 
 constructed by method (v), HPS 
 that attained the median  test $\mathrm{R}^2$ in Table~\ref{table:phase1}.
 To solve an MILP with the formulation, we used 
{\tt  CPLEX} version 12.10.
Tables~\ref{table:stages_4_5_At} and \ref{table:stages_4_5_FlmL}  show
   the computational results of the experiment
in this stage for the two properties, 
 where we denote the following:
\begin{enumerate} [nosep,  leftmargin=*]
  
\item[-]  
$n_\LB$: a lower bound on the number of non-hydrogen atoms 
in  a chemical graph $\Co$ to be inferred; 

\item[-]  
  $ \underline{y}^*,~\overline{y}^* $:  
 lower and upper bounds $\underline{y}^*, \overline{y}^*\in \R$ 
  on the value $a(\Co)$ of a chemical graph $\Co$ to be inferred; 
 For {\sc At}, we use range of the original values $a(\C)$ before normalization.
 For {\sc FlmL}, we use the logarithmic scale $\log a(\C)$ as the range of target values.
 
\item[-]  
 $\#$v (resp.,  $\#$c): 
 the number  of variables (resp., constraints)  in the MILP;  
  
\item[-]   
 I-time: the   time (sec.) to solve the MILP; 

\item[-]  
    $n$:  the number  $n(\Co^\dagger)$  of  non-hydrogen atoms
     in the chemical graph $\Co^\dagger$   inferred by solving the MILP;   
     
\item[-]  
  $\nint$:  the number  $\nint(\Co^\dagger)$ of interior-vertices in
  the inferred chemical graph $\Co^\dagger$; and 
      
\item[-]  
$\eta$: the predicted property value 
$\eta(f(\Co^\dagger))$ of the inferred chemical graph $\Co^\dagger$.
\end{enumerate}

\begin{table}[h!]\caption{Results of inferring a chemical graph $\Co^\dagger$ 
and generating recombination solutions for  {\sc At} with $\Lambda_3$.} 
 \begin{center}
 \begin{tabular}{@{}  c  c   c  r r c r r c  r r r     @{}}\hline                
 inst. & $n_\LB$ &  $ \underline{y}^*,~\overline{y}^* $ & $\#$v~  &  $\#$c~   &  
   {\small I-time}  & $n$~  &  \!\!$\nint$  &  $\eta $ \!\! & 
                 {\small  D-time} &  {\small $\Co$-LB} &  {\small $\#\Co$}    \\ \hline
  $I_{\mathrm{a}}$ & 30 & 240, 250  & 9846  & 9250 &  3.61 &  44 &  25  & 241.98  & 0.111 &  12  & 12 \\
  $I_{\mathrm{b}}^1$ &35  & 240, 250  & 10428 &  6904  & 3.07 &  37  & 15 &  247.71 &  0.0481 &  8 &  8\\
  $I_{\mathrm{b}}^2$ &  45  & 240, 250  & 13113 &  10019 &  12.8 &  50  & 25  & 241.51  & 0.155  & 432  & 100\\
 $I_{\mathrm{b}}^3$ &  45 &  190, 200 &  12909 &  10022 &  10.3 &  50 &  25 &  197.34  & 0.267 &  208  & 100\\
 $I_{\mathrm{b}}^4$ &  45 &  300, 310 &  12705 &  10025 &  12.1 &  48 &  29 &  300.18 &  0.192 &  432 &  100\\
$I_{\mathrm{c}}$  & 50  & 360, 370 &  7875 &  8734 &  1.29 &  50 &  33  & 361.62 &  0.0163 &  1 &  1\\
 $I_{\mathrm{d}}$  &  40 &  230, 240 &  5479 &  6775 & 2.88 &  43 &  23  & 237.72 &  0.184  & 10496 &  100 \\
   \hline
   \end{tabular}\end{center}\label{table:stages_4_5_At}
\end{table}

 \begin{table}[h!]\caption{ Results of inferring a chemical graph $\Co^\dagger$
 and generating recombination solutions  for {\sc FlmL} with $\Lambda_6$.}  
 \begin{center}
 \begin{tabular}{@{}  c  c   c  r r r r r c  r r r     @{}}\hline                
 inst. & $n_\LB$ &  $ ~\underline{y}^*,~\overline{y}^* $ & $\#$v~  &  $\#$c~   &  
   {\small I-time}  & $n$~  &  \!\!$\nint$  &  $\eta $ \!\! & 
                 {\small  D-time} &  {\small $\Co$-LB} &  {\small $\#\Co$}    \\ \hline
   $I_{\mathrm{a}}$ & 30 & -0.55, -0.5 & 11794 & 12922 & 4.17 & 38 & 23 & -0.551 & 0.0692 & 1 & 1\\
$I_{\mathrm{b}}^1$ & 35 & -0.15, -0.1 & 11107 & 10423 & 2.82 & 35 & 9 & -0.140 & 0.0163 & 2 & 2\\
 $I_{\mathrm{b}}^2$ & 45 & ~~-0.5,~-0.45 & 13426 & 13871 & 28.6 & 50 & 28 & -0.494 & 12.8 & 207594 & 100\\
$I_{\mathrm{b}}^3$ & 45 & ~~-0.5, -0.45 & 12974 & 13535 & 20.2 & 45 & 25 & -0.483 & 17.2 & 2215023 & 100\\
$I_{\mathrm{b}}^4$ & 45 & -0.45, -0.4 & 12746 & 13534 & 12.5 & 46 & 27 & -0.445 & 0.0889 & 5040 & 100\\
 $I_{\mathrm{c}}$   & 50 & -0.55, -0.5 & 9386 & 11000 & 0.878 & 50 & 33 & -0.509 & 0.0162 & 1 & 1\\
  $I_{\mathrm{d}}$   & 40 & ~~~0.2, 0.25 & 6184 & 7821 & 9.18 & 44 & 23 & 0.204 & 0.16 & 21600 & 100\\
   \hline   
   \end{tabular}\end{center}\label{table:stages_4_5_FlmL}
\end{table}

\begin{figure}[!htb]
\begin{center} 
\includegraphics[width=.98\columnwidth]{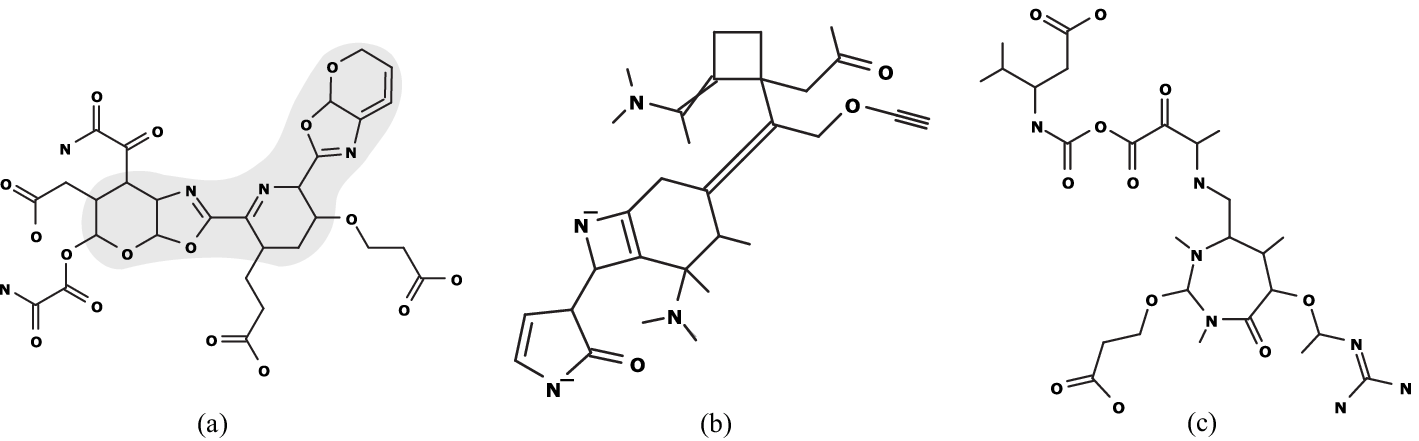}
\end{center}
\caption{ 
}
\label{fig:MILP_solutions}  
\end{figure}

 Figure~\ref{fig:MILP_solutions}(a) illustrates  the chemical graph  $\Co^\dagger$  inferred
 from   $I_{\mathrm{c}}$ with $(\underline{y}^*, \overline{y}^*) =(360, 370)$   of  {\sc At}
  in Table~\ref{table:stages_4_5_At}.  
 
Figure~\ref{fig:MILP_solutions}(b) 
(resp.,  Figure~\ref{fig:MILP_solutions}(c)) illustrates  the chemical graph  $\Co^\dagger$  inferred 
 from  $I_{\mathrm{a}}$ with $(\underline{y}^*, \overline{y}^*) =(-0.55, -0.5)$  
 (resp.,  $I_{\mathrm{d}}$ with $(\underline{y}^*, \overline{y}^*) =(0.2, 0.25)$) 
 of  {\sc  FlmL}   in Table~\ref{table:stages_4_5_FlmL}.

From Tables~\ref{table:stages_4_5_At} and \ref{table:stages_4_5_FlmL},
we observe that an instance with a large number of variables and constraints 
takes more running time than those with a smaller size in general.
All instances in this experiment are solved in a few seconds to around 30 seconds
with our MILP formulation. 


\medskip \noindent
{\bf Generating Recombination Solutions } 
Let   $\Co^\dagger$ be a chemical graph obtained by solving
the MILP $\mathcal{M}_{f,\eta,\sigma}$ for the inverse problem.
We here execute  a stage of generating recombination solutions
 $\C^*\in \mathcal{G}_\sigma$ of $\C^\dagger$
 such that $f(\C^*)=x^*=f(\C^\dagger)$.
 
We execute an  algorithm for generating chemical isomers of   $\Co^\dagger$
up to 100 when the number of all chemical isomers exceeds 100.
For this, we use a dynamic programming algorithm~\cite{ZAHZNA21}. 
The algorithm first decomposes $\Co^\dagger$ into a set of acyclic chemical graphs,
next replaces each acyclic chemical graph $T$ with another  acyclic chemical graph $T'$ that admits
the same feature vector as that of $T$
 and finally assembles the resulting acyclic chemical graphs
into a chemical isomer $\Co^*$ of $\Co^\dagger$. 
The algorithm can compute a lower bound 
on the total number of all chemical isomers $\Co^\dagger$
without generating all of them.

Tables~\ref{table:stages_4_5_At} and \ref{table:stages_4_5_FlmL}  show
   the computational results of the experiment
in this stage  for the two properties $\pi\in \{${\sc At,  FlmL}$\}$, 
 where we denote the following:
\begin{enumerate}[nosep,  leftmargin=*]
\item[-]
 D-time: the running time (sec.) to execute the dynamic programming algorithm
  to compute a lower bound on the number 
 of all chemical isomers  $\Co^*$ of  $\Co^\dagger$   
 and generate all (or up to 100) chemical isomers $\Co^*$;
 
\item[-]
 $\Co$-LB: a lower bound on the number of all chemical isomers $\Co^*$ of 
$\C^\dagger$;~and

\item[-]
 $\#\Co$: the number of all (or up to 100) chemical isomers $\Co^*$ of  $\Co^\dagger$  
 generated in this stage.
\end{enumerate} 
  
From Tables~\ref{table:stages_4_5_At} and \ref{table:stages_4_5_FlmL}, we observe  
 the running time and the number of generated recombination solutions  in this stage. 

For the tested properties, 
the chemical graph $\Co^\dagger$   in  $I_{\mathrm{b}}^2$, $I_{\mathrm{b}}^3$
and $I_{\mathrm{d}}$ admits a large number of 
chemical isomers $\Co^*$, 
where a lower bound $\Co$-LB  on the number of chemical isomers
is derived without generating all of them.  
The running time
 for computing the lower bound and generating up to 100 target chemical graphs  
 is at most  18 second. 
 For some chemical graphs $\Co^\dagger$, the number of  chemical isomers found by our algorithm was small.
 This is because some of acyclic chemical graphs in the decomposition of $\Co^\dagger$
 has no alternative acyclic chemical graph other than the original one. 


\medskip \noindent
{\bf Generating Neighbor Solutions  } 
Let   $\Co^\dagger$ be a chemical graph obtained by solving
the MILP $\mathcal{M}_{f,\eta,\sigma}$ for the inverse problem.
We executed a stage of generating neighbor solutions 
 of   $\Co^\dagger$.
 
We select an MILP for the inverse problem with a prediction function $\eta$
 such that a solution $\C^\dagger$ of the MILP 
admits only two isomers $\C^*$ in the stage of
generating recombination solutions;
i.e.,  instance 
$I_{\mathrm{c}}$ for property   {\sc At} with $\Lambda_3$ and  
 instances  $I_{\mathrm{a}}$,  $I_{\mathrm{b}}^4$ and $I_{\mathrm{c}}$ 
for property   {\sc FlmL}  with $\Lambda_6$.

In this experiment, we add to the MILP $\mathcal{M}_{f,\eta,\sigma}$
  an additional set $\Theta$ of two linear constraints on linear and quadratic descriptors
as follows.
For the two constraints, we use the prediction functions $\eta_{\pi}$ constructed
by RLR for properties  $\pi\in \{${\sc Mp},  {\sc Sl}$\}$ with $\Lambda_3$
in Table~\ref{table:phase1}.

We regard each of  $\eta_{\mbox{\tiny{\sc Mp}}}$ and  $\eta_{\mbox{\tiny{\sc Sl}}}$
  as  a function from  $\R^{|D^\mathrm{union}_{\pi}|}$ to $\R$ for $\pi\in\{${\sc At}, {\sc FlmL}$\}$.
We set $p_{\mathrm{dim}}:=2$ and let  $\Theta$ consist of two 
 linear constraints  $\theta_1:=\eta_{\mbox{\tiny{\sc Mp}}}$
  and $\theta_2:=\eta_{\mbox{\tiny{\sc Sl}}}$.
We select   $\delta\in \{ 0.01, 0.05, 0.1, 0.15\}$  which defines a two-dimensional grid space 
where  $\C^\dagger$ is mapped to the origin
(see \cite{gridAZHZNA21} for the detail on the neighbors). 
We choose a set $N_0$ of 48 neighbors 
of the origin  $\C^\dagger$ in the grid search space. 
%
For each instance, we check the feasibility of neighbors in $N_0$ 
 in a non-decreasing order of the distance between
 the neighbor and the origin.
For each feasible neighbor $z \in N_0$, output a feasible solution $\C^\dagger_z$
of  the augmented MILP instance. 
We set a time limit for checking the feasibility of a neighbor to be 300 seconds, 
and we skip a neighbor when the corresponding MILP is not solved within the time limit.
We also ignore any neighbor $z \in N_0$
without testing the feasibility of $z$ if we find an infeasible neighbor $z'\in N_0$
such that $z'$ is closer to the origin than $z$ is.

Table~\ref{table:neighbor}    shows
   the computational results of the experiment for the three instances, 
 where we denote the following:
\begin{enumerate}[nosep,  leftmargin=*]
\item[-]
(inst., $\pi$):   topological specification $I$ and property $\pi$;

\item[-] $n$: the number of non-hydrogen atoms in the tested instance;  

\item[-] $\delta$: the size of a sub-region in the grid search space;

\item[-] 
\#sol:   the number of new chemical graphs
obtained from the neighbor set $N_0$; 

\item[-] 
\#infs: the number of   neighbors in $N_0$
that are found to be infeasible during the search procedure; 

\item[-] 
\#ign:  the number of   neighbors  in $N_0$
that are ignored during the search procedure; 

\item[-] 
\#TO: the number of  neighbors in $N_0$
such that the time for feasibility check exceeds
the time limit of 300 seconds during the search procedure.

\end{enumerate}

 \begin{table}[h!]\caption{Results of generating neighbor solutions of $\C^\dagger$.  } 
 \begin{center}
 \begin{tabular}{@{}  c   r r r    r r r      @{}}\hline             
(inst., $\pi$) & $n$ & $\delta$~~ &  \#sol  &  \#infs & \#ign  & \#TO   \\ \hline 
   ($I_{\mathrm{c}}$,{\sc At})   &  50 & 0.01 & 9  & 0 & 0 & 39     \\ 
  ($I_{\mathrm{a}}$,{\sc FlmL})    & 30  &0.05 & 41   & 0 & 0 & 7   \\ 
  ($I_{\mathrm{b}}^4$,{\sc FlmL}) & 45  &0.15 & 38    & 0  & 0 & 10     \\ 
  ($I_{\mathrm{c}}$,{\sc FlmL})   &  40 & 0.05 & 15  & 0 & 0 & 33     \\  
   \hline
   \end{tabular}\end{center}\label{table:neighbor}  
\end{table}
 
In many solvers such as CPLEX, an MILP is solved by
an algorithm based on the branch-and-bound method, which 
sometimes takes 
an extremely large execution time for the same size of instances. 
We introduce a time limit to bound a running time
of testing the feasibility of neighbors in   $N_0$  to skip such instances. 
From Table~\ref{table:neighbor}, we observe
 that   some number of neighbor solutions of   $\C^\dagger$
 could be successfully generated for each of the four instances.


\section{Concluding Remarks}\label{sec:conclude}
 
 In the framework of inferring chemical graphs,  
 the descriptors of a prediction function were mainly defined
to be the frequencies of local graph structures 
 in the two-layered model and defining descriptors in such a way is 
 important to derive a compact MILP formulation
 in the second phase of the framework .
To improve the performance of prediction functions based on the same definition of descriptors,
this paper  proposed a method of splitting a given data set into two subsets
by a hyperplane in the feature space so that the first and second subsets
mainly consist of compounds with observed values lower and higher than a threshold,
respectively. 
A prediction function is obtained by combining prediction functions for the first and second subsets
constructed independently, where the hyperplane is used to decide which of the two prediction functions
is applied for a given feature vector. 
Our experimental results show that the proposed method improved
the learning performance of chemical properties such as flammable limits and odor threshold and that the MILP in the second phase is solvable
for instances of inferring a chemical graph with around 50 non-hydrogen atoms. 
 It is left as a future work to extend our new method of splitting a data set
 so that a given data set is repeatedly split into smaller subsets
with a narrower range of observed values 
when the size of the data set is large enough.


\clearpage
 \appendix
\centerline{\bf\LARGE Appendix}

\section{A Full Description of Descriptors}\label{sec:descriptor}

Associated with the two functions 
$\alpha$ and $\beta$ in a chemical graph $\Co=(H,\alpha,\beta)$,
we introduce   functions  
 $\ac: V(E)\to (\Lambda\setminus\{\ttH\})\times (\Lambda\setminus\{\ttH\})\times [1,3]$, 
 $\cs: V(E)\to (\Lambda\setminus\{\ttH\})\times [1,6]$ and
$\ec: V(E)\to ((\Lambda\setminus\{\ttH\})\times [1,6])\times ((\Lambda\setminus\{\ttH\})\times [1,6])\times [1,3]$
in the following. 

 To represent  a feature of the exterior  of  $\Co$, 
  a  chemical rooted tree in $\mathcal{T}(\Co)$ is
  called a {\em fringe-configuration} of $\Co$. 

We also represent leaf-edges in the exterior of $\Co$.
For a leaf-edge $uv\in E(\anC)$ with $\deg_{\anC}(u)=1$, we define
the {\em adjacency-configuration} of $e$ to be an ordered tuple
$(\alpha(u),\alpha(v),\beta(uv))$. 
Define 
\[ \Gac^\lf\triangleq \{(\ta,\tb,m)\mid \ta,\tb\in\Lambda, 
m\in[1,\min\{\val(\ta),\val(\tb)\}]\} \]
as a set of possible adjacency-configurations for leaf-edges. 

To  represent a feature of an interior-vertex $v\in V^\inte(\Co)$ such that
$\alpha(v)=\ta$  and  $\deg_{\anC}(v)=d$
(i.e., the number of non-hydrogen atoms adjacent to $v$ is $d$) 
   in a chemical   graph  $\Co=(H,\alpha,\beta)$,
 we use  a pair $(\ta, d)\in (\Lambda\setminus\{{\tt H}\})\times [1,4]$,
 which we call the {\em chemical symbol} $\cs(v)$ of the vertex $v$.
 We treat $(\ta, d)$ as a single symbol $\ta d$,  and  
define $\Ldg$   to be  the set of all chemical symbols
$\mu=\ta d\in  (\Lambda\setminus\{{\tt H}\})\times [1,4]$.  

We define a method for featuring interior-edges  as follows.
Let $e=uv\in E^\inte(\Co)$  be 
 an interior-edge $e=uv\in E^\inte(\Co)$ 
 such that $\alpha(u)=\ta$, $\alpha(v)=\tb$ and $\beta(e)=m$ 
   in a chemical graph  $\Co=(H,\alpha,\beta)$.
To feature this edge $e$, 
 we use a tuple $(\ta,\tb,m)\in (\Lambda\setminus\{{\tt H}\})
    \times (\Lambda\setminus\{{\tt H}\})\times [1,3]$,
 which we call the {\em adjacency-configuration} $\ac(e)$ of the edge $e$. 
 We introduce a total order $<$ over the elements in $\Lambda$
 to distinguish  between $(\ta,\tb, m)$ and $(\tb,\ta, m)$ 
 $(\ta\neq \tb)$ notationally.
 For a tuple  $\nu=(\ta,\tb, m)$,
 let $\overline{\nu}$ denote the tuple $(\tb,\ta, m)$.

Let $e=uv\in E^\inte(\Co)$  be 
an  interior-edge $e=uv\in E^\inte(\Co)$ 
 such that $\cs(u)=\mu$, $\cs(v)=\mu'$ and $\beta(e)=m$ 
   in a chemical  graph  $\Co=(H,\alpha,\beta)$.
To feature this edge $e$, 
 we use a tuple $(\mu,\mu',m)\in \Ldg\times \Ldg\times [1,3]$, 
 which we call  the {\em edge-configuration} $\ec(e)$ of the edge $e$. 
 We introduce a total order $<$ over the elements in $\Ldg$
 to distinguish between $(\mu,\mu', m)$ and $(\mu', \mu, m)$ 
 $(\mu \neq \mu')$ notationally. 
 For a tuple  $\gamma=(\mu,\mu', m)$,
 let $\overline{\gamma}$ denote the tuple $(\mu', \mu, m)$. 

Let $\pi$ be a chemical property for which we will construct
a prediction function $\eta$ from a feature
vector $f(\Co)$ of a chemical graph $\Co$ 
to a predicted value $y\in \mathbb{R}$
for the  chemical property of $\Co$.

We first choose a set $\Lambda$ of chemical elements
 and then collect a data set  $\mathcal{C}_{\pi}$ of
  chemical compounds  $C$ whose 
  chemical elements belong to $\Lambda$,
  where we regard  $\mathcal{C}_{\pi}$ as a set of chemical graphs $\Co$
  that represent the chemical compounds $C$  in  $\mathcal{C}_{\pi}$.
To define the interior/exterior of 
chemical graphs  $\Co\in \mathcal{C}_{\pi}$,
we  next choose a branch-parameter ${\rho}$, where
 we recommend ${\rho}=2$.  
 
Let $\Lambda^\inte(\mathcal{C}_\pi)\subseteq \Lambda$ 
(resp., $\Lambda^\ex(\mathcal{C}_\pi)\subseteq \Lambda$)
denote the set  of chemical elements  used in
the set $V^\inte(\Co)$ of interior-vertices
(resp., the set $V^\ex(\Co)$ of  exterior-vertices) of $\Co$
 over all chemical graphs $\Co\in \mathcal{C}_\pi$, 
and $\Gamma^\inte(\mathcal{C}_\pi)$
denote the set of edge-configurations used in
the set $E^\inte(\Co)$  of interior-edges in $\Co$
 over all chemical graphs $\Co\in \mathcal{C}_\pi$. 
Let $\mathcal{F}(\mathcal{C}_\pi)$ denote the set of
chemical rooted trees $\psi$  
r-isomorphic to a chemical rooted tree in $\mathcal{T}(\Co)$
  over all chemical graphs $\Co\in \mathcal{C}_\pi$,
  where possibly a chemical rooted tree $\psi\in \mathcal{F}(\mathcal{C}_\pi)$
  consists of a single chemical element $\ta\in \Lambda\setminus \{{\tt H}\}$.
  
We define an integer encoding of a finite set $A$ of elements
to be a bijection $\sigma: A \to [1, |A|]$, 
where we denote by $[A]$   the set $[1, |A|]$ of integers.
Introduce  an integer coding of each of the   sets 
$\Lambda^\inte(\mathcal{C}_\pi)$, $\Lambda^\ex(\mathcal{C}_\pi)$, 
$\Gamma^\inte(\mathcal{C}_\pi)$ and $\mathcal{F}(\mathcal{C}_\pi)$. 
Let $[\ta]^\inte$  
(resp., $[\ta]^\ex$)  denote   
the coded integer of  an element $\ta\in \Lambda^\inte(\mathcal{C}_\pi)$
(resp., $\ta\in \Lambda^\ex(\mathcal{C}_\pi)$),  
$[\gamma]$   denote  
the coded integer of  an element $\gamma$ in $\Gamma^\inte(\mathcal{C}_\pi)$
and 
$[\psi]$   denote  an element $\psi$ in $\mathcal{F}(\mathcal{C}_\pi)$. 
 
 Over 99\% of  chemical compounds $\Co$ with up to
  100 non-hydrogen atoms in  PubChem  have degree at most 4
 in the hydrogen-suppressed graph $\anC$~\cite{AZSSSZNA20}. 
We assume that a chemical graph $\Co$
 treated in this paper satisfies  $\deg_{\anC}(v)\leq 4$
in the hydrogen-suppressed graph $\anC$.
 
In our model, we  use an integer 
  $\mathrm{mass}^*(\ta)=\lfloor 10\cdot \mathrm{mass}(\ta)\rfloor$, 
 for each $\ta\in \Lambda$.
 
 For a chemical property $\pi$,
 we define a set $D_\pi^{(1)}$ of descriptors  
  of a  chemical graph $\C=(H,\alpha,\beta)\in \mathcal{C}_{\pi}$ 
  to be  the following  
non-negative integers $\dcp_i(\C)$, $i\in [1,K_1]$, where 
$K_1= 14+ |\Lambda^\inte(\mathcal{C}_\pi)|+|\Lambda^\ex(\mathcal{C}_\pi)|
         +|\Gamma^\inte(\mathcal{C}_\pi)|+|\mathcal{F}(\mathcal{C}_\pi)|+|\Gac^\lf|$. 


\begin{enumerate}  
\item   
$\dcp_1(\C)$: the number  $|V(H)|-|\VH|$ of non-hydrogen atoms  in  $\C$.  
 
\item   
$\dcp_2(\C)$: the rank 
of   $\C$ (i.e., the minimum number of edges to be removed to make the graph 
acyclic).  

\item 
$\dcp_3(\C)$:  the number $|V^\inte(\C)|$ of interior-vertices in  $\C$.
  
\item 
$\dcp_4(\C)$: 
the average $\overline{\mathrm{ms}}(\C)$ of mass$^*$ 
over all atoms in $\C$; \\
 i.e., $\overline{\mathrm{ms}}(\C)\triangleq 
 \frac{1}{|V(H)|}\sum_{v\in V(H)}\mathrm{mass}^*(\alpha(v))$. 

\item 
$\dcp_i(\C)$,  $i=4+d,   d\in [1,4]$: 
the number $\dg_d^{\oH} (\C)$ 
 of non-hydrogen vertices $v\in V(H)\setminus \VH$
 of degree $\deg_{\anC}(v)=d$
 in the hydrogen-suppressed chemical graph $\anC$.  
 
\item 
$\dcp_i(\C)$,  $i=8+d,   d\in [1,4]$: 
the number $\dg_d^\inte(\C)$
 of interior-vertices of interior-degree  $\deg_{\C^\inte}(v)=d$
  in the interior $\C^\inte=(V^\inte(\C),E^\inte(\C))$ of  $\C$. 
  
   
\item $\dcp_i(\C)$, $i=12+m$,  $m\in[2,3]$: 
the number $\bd_m^\inte(\C)$
 of  interior-edges with bond multiplicity $m$ in  $\C$; 
 i.e., $\bd_m^\inte(\C)\triangleq |\{e\in E^\inte(\C)\mid \beta(e)=m\}|$.

\item $\dcp_i(\C)$, $i=14+[\ta]^\inte$, 
 $\ta\in \Lambda^\inte(\mathcal{C}_\pi)$: 
 the frequency $\na_\ta^\inte(\C)=|V_\ta(\C)\cap V^\inte(\C) |$ 
 of chemical element $\ta$ in
 the set $V^\inte(\C)$ of  interior-vertices in  $\C$. 
 
\item $\dcp_i(\C)$, 
$i=14+|\Lambda^\inte(\mathcal{C}_\pi)|+[\ta]^\ex$, 
 $\ta\in \Lambda^\ex(\mathcal{C}_\pi)$: 
 the frequency $\na_\ta^\ex(\C)=|V_\ta(\C)\cap V^\ex(\C) |$
  of chemical element $\ta$ in
 the set $V^\ex(\C)$ of  exterior-vertices in  $\C$. 
 
\item $\dcp_i(\C)$, 
$i=14+|\Lambda^\inte(\mathcal{C}_\pi)|+|\Lambda^\ex(\mathcal{C}_\pi)|+ [\gamma]$, 
$\gamma \in \Gamma^\inte(\mathcal{C}_\pi)$: 
the frequency $\ec_{\gamma} (\Co)$ of edge-configuration $\gamma$
in the set $E^\inte(\C)$ of interior-edges   in  $\C$.

\item $\dcp_i(\C)$, 
$i= 14+|\Lambda^\inte(\mathcal{C}_\pi)|+|\Lambda^\ex(\mathcal{C}_\pi)|
+ |\Gamma^\inte(\mathcal{C}_\pi)|+ [\psi]$,  
 $\psi \in \mathcal{F}(\mathcal{C}_\pi)$: 
the frequency $\fc_{\psi}(\C)$ of fringe-configuration $\psi $
in the set of ${\rho}$-fringe-trees in  $\C$. 

\item $\dcp_i(\C)$, 
$i= 14+|\Lambda^\inte(\mathcal{C}_\pi)|+|\Lambda^\ex(\mathcal{C}_\pi)|
+ |\Gamma^\inte(\mathcal{C}_\pi)|+|\mathcal{F}(\mathcal{C}_\pi)|+ [\nu]$,  
 $\nu \in \Gac^\lf$: 
the frequency $\ac_{\nu}^\lf(\C)$ of adjacency-configuration $\nu$
in the set of leaf-edges in  $\anC$. 
\end{enumerate} 


In this paper, we also use a method of generating quadratic descriptors.
For this, we first normalize each descriptor  $\dcp_i(\C), i\in[1,K_1]$
to a value $x(i)$ between 0 and 1 by scaling the minimum and   maximum
values to 0 and 1, respectively.
Then construct a set  $D_\pi^{(2)}:=\{x(i)x(j)\mid 1\leq i\leq j\leq K_1\}\cup  \{ x(i)(1-x(j))\mid  i,j\in [1,K_1]\}$ of  $(3K_1^2+K_1)/2$ quadratic descriptors.
Then we reduce the union $D_\pi^{(1)}\cup D_\pi^{(2)}$
 to a subset to construct a prediction function by a procedure  
 proposed by Zhu~et al.~\cite{Q_ZACIHZNA22}.


 
\section{Specifying Target Chemical Graphs}\label{sec:specification} 

Given a prediction function $\eta$ and 
a target value $y^*\in \mathbb{R}$, 
we call a chemical graph $\C^*$ such that $\eta(x^*)=y^*$
for the feature vector $x^*=f(\C^*)$ a {\em target chemical graph}.
This section  presents a set of rules for 
 specifying  topological substructure
  of a target chemical graph in a flexible way in the second phase of the framework.

We first describe how to reduce a chemical graph $\C=(H,\alpha,\beta)$ into
an abstract form based on which our specification rules will be defined.
To illustrate the reduction process,
we use the chemical graph $\C=(H,\alpha,\beta)$
such that $\anC$ is given in Figure~\ref{fig:example_chemical_graph}.
 
\begin{figure}[h!] \begin{center}
\includegraphics[width=.65\columnwidth]{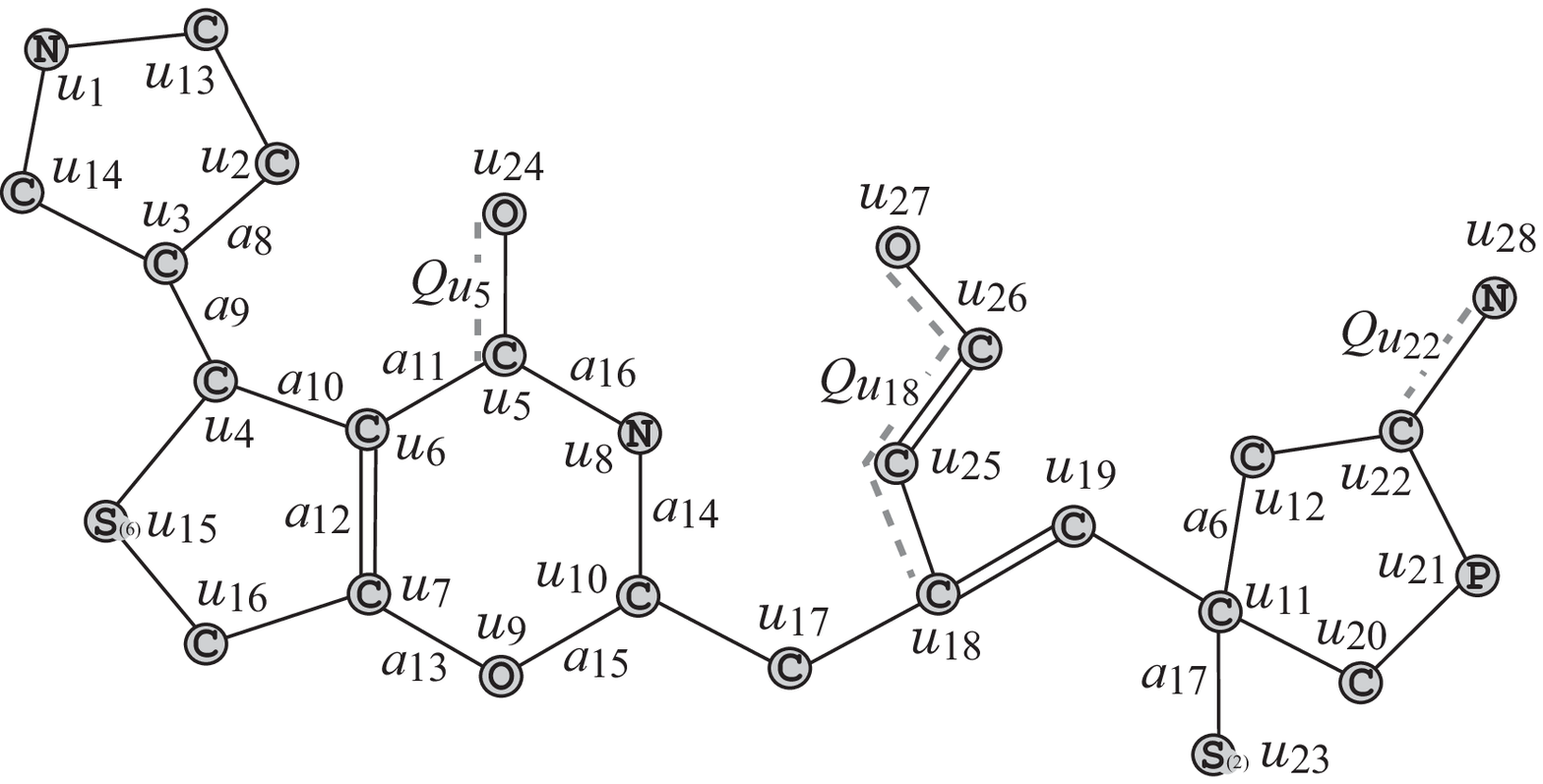}
\end{center} \caption{The interior $H^\inte$ of
chemical graph $\C$ with $\anC$ 
  in Figure~\ref{fig:example_chemical_graph} for ${\rho}=2$.
}
\label{fig:specification_example_interior} \end{figure}

\begin{figure}[h!] \begin{center}
\includegraphics[width=.98\columnwidth]{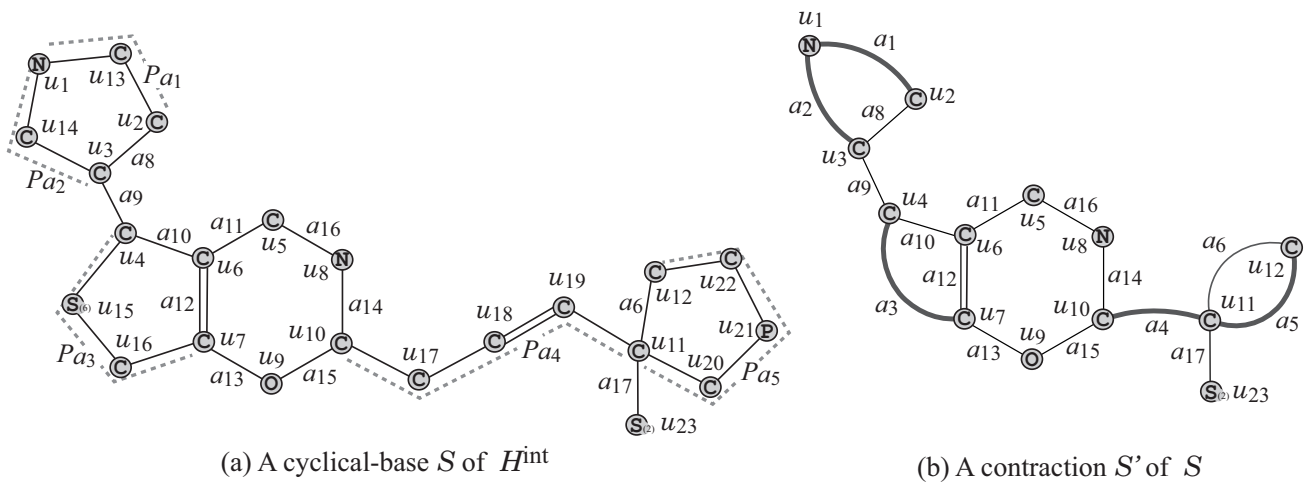}
\end{center} \caption{
(a) A cyclical-base  
$S=H^\inte- \bigcup_{u\in \{u_5,u_{18},u_{22}\}}(V(Q_u)\setminus \{u\})$
of the interior  $H^\inte$ in Figure~\ref{fig:specification_example_interior};
(b) A contraction $S'$ of  $S$ for a pure path set 
 $\mathcal{P}=\{P_{a_1},P_{a_2},\ldots,P_{a_5}\}$ 
in (a),
where a new edge obtained by contracting a pure path is depicted
with a thick line.   
}
\label{fig:specification_example_R2_3} \end{figure}

 \begin{enumerate}
 \item[R1] {\bf Removal of all ${\rho}$-fringe-trees: } 
The interior $H^\inte=(V^\inte(\C),E^\inte(\C))$ of $\C$ 
is obtained by removing the non-root vertices of 
each ${\rho}$-fringe-trees $\C[u]\in\mathcal{T}(\C), u\in V^\inte(\C)$. 
Figure~\ref{fig:specification_example_interior} illustrates
the interior $H^\inte$ of 
chemical graph $\C$ with ${\rho}=2$
  in Figure~\ref{fig:example_chemical_graph}. 
  
 \item[R2] {\bf Removal of some leaf paths: } 
 We call a $u,v$-path $Q$ in $H^\inte$  a {\em leaf path} if 
  vertex $v$ is a leaf-vertex of $H^\inte$
  and the degree of each internal vertex of $Q$  in $H^\inte$  is 2,
  where we regard that $Q$ is rooted at vertex $u$. 
A connected subgraph $S$ of the interior $H^\inte$ of $\C$  
is called a {\em cyclical-base}
if $S$ is obtained from $H$
by removing the vertices in $V(Q_u)\setminus \{u \}, u\in X$ 
for a subset $X$ of interior-vertices  and a set  $\{Q_u \mid u\in X\}$ of leaf 
 $u,v$-paths $Q_u$  such that    
 no two paths $Q_u$ and $Q_{u'}$ share a vertex.
Figure~\ref{fig:specification_example_R2_3}(a) illustrates
a cyclical-base  $S=H^\inte- \bigcup_{u\in X}(V(Q_u)\setminus \{u\})$
of the interior  $H^\inte$  
for a set 
$\{Q_{u_5}=(u_5,u_{24}), 
     Q_{u_{18}}=(u_{18},u_{25},u_{26},u_{27}),
     Q_{u_{22}}=(u_{22},u_{28})\}$ of leaf  paths 
in Figure~\ref{fig:specification_example_interior}.  

 \item[R3] {\bf Contraction of some pure paths: } 
 A path in $S$ is called {\em pure} 
 if  each internal vertex of the path  is of degree 2. 
 Choose a set $\mathcal{P}$ of several pure paths in $S$ 
 so that no two paths share  vertices except for their end-vertices. 
 A graph $S'$ is called a {\em contraction} of a graph $S$
  (with respect to $\mathcal{P}$) 
 if $S'$ is obtained from $S$ by replacing 
 each pure $u,v$-path  with a single edge $a=uv$,
 where $S'$ may contain multiple edges between the same pair of adjacent vertices.
Figure~\ref{fig:specification_example_R2_3}(b) illustrates
a contraction $S'$ obtained from 
the chemical graph  $S$
by contracting each $uv$-path $P_a\in  \mathcal{P}$ into a new edge $a=uv$,
where $a_1=u_1 u_{2},  a_2=u_1 u_{3},  a_3=u_4 u_{7}, a_4=u_{10}u_{11}$
and $a_5=u_{11}u_{12}$ and 
 $\mathcal{P}=\{
 P_{a_1}=(u_1,u_{13},u_{2}), 
 P_{a_2}=(u_{1},u_{14},u_{3}),
 P_{a_3}=(u_{4},u_{15},u_{16},u_{7}), 
 P_{a_4}=(u_{10},u_{17},u_{18},u_{19},u_{11}),
 P_{a_5}=(u_{11},u_{20},u_{21},u_{22},u_{12})\}$ of pure paths 
in Figure~\ref{fig:specification_example_R2_3}(a). 
\end{enumerate}

We will define a set of rules so that 
a chemical graph can be obtained 
from a graph (called a seed graph in the next section)
by applying processes R3 to R1 in a reverse way. 
We specify topological substructures of a target chemical graph
with a tuple  $(\GC,\sint,\sce)$  called  a {\em target specification}
defined under the set of the following rules. 

\subsection*{Seed Graph}

A  {\em seed graph} $\GC=(\VC,\EC)$ is defined
to be a graph (possibly with multiple edges) such that 
the edge set $\EC$ consists of four sets 
$\Et$, $\Ew$, $\Ez$ and $\Eew$, 
where each of them can be empty.
A seed graph plays a role of the most abstract form $S'$ in R3.  
Figure~\ref{fig:specification_example_1}(a) illustrates an example of a seed graph
$\GC$,  
where $\VC=\{u_1,u_2,\ldots,u_{12},u_{23}\}$, 
$\Et=\{a_1,a_2,\ldots,a_5\}$, 
$\Ew=\{a_6\}$,
$\Ez=\{a_7\}$ and 
$\Eew=\{a_8,a_9,\ldots,a_{16}\}$.

 A {\em subdivision} $S$ of $\GC$  
is a graph constructed from a seed graph $\GC$ 
according to the following rules:
\begin{enumerate}[leftmargin=*]
\item[-]
Each edge $e=uv\in \Et$ is replaced
with a $u,v$-path $P_e$ of length at least 2;

\item[-] 
Each edge $e=uv\in \Ew$ is replaced
with a $u,v$-path $P_e$ of length at least 1
(equivalently $e$ is directly used or replaced with
a $u,v$-path $P_e$ of length at least 2);

\item[-] 
Each edge $e\in \Ez$ is either used or discarded, where 
 $\Ez$ is required to be chosen as a non-separating edge subset of
 $E(\GC)$ since otherwise the connectivity of a final chemical graph $\Co$
 is not guaranteed; 
and 

\item[-]
Each edge $e\in \Eew$ is always used directly. 
\end{enumerate}

We allow a possible elimination of edges in $\Ez$ as an optional rule
in constructing a target chemical graph from a seed graph, 
even though such an operation has 
not been included in the process R3. 
A subdivision  $S$ plays a role of a cyclical-base in R2. 
A target chemical graph $\C=(H,\alpha,\beta)$ will contain  $S$  as a subgraph
of the interior $H^\inte$ of $\C$.


\subsection*{Interior-specification}

A graph $H^*$ that serves as the interior $H^\inte$ of
a target chemical graph $\C$ will be constructed as follows.
First construct a subdivision  $S$ of a seed graph $\GC$ 
by replacing each edge $e=u u'\in \Et\cup\Ew$
with a pure $u,u'$-path $P_e$.
Next construct a supergraph $H^*$ of $S$ by 
attaching a leaf path $Q_v$ at each vertex $v\in \VC$ or
at an internal vertex $v\in V(P_e)\setminus\{u,u'\}$ 
of each pure $u,u'$-path $P_e$ for some edge $e=uu'\in \Et\cup\Ew$,
where possibly $Q_v=(v), E(Q_v)=\emptyset$ 
(i.e., we do not attach any new edges to $v$).
We introduce the following rules for specifying
 the size of $H^*$, the length $|E(P_e)|$  of
a pure path  $P_e$,  the length $|E(Q_v)|$ of
a   leaf path $Q_v$, the number of  leaf paths $Q_v$
and a bond-multiplicity of each interior-edge,
where we call the set of prescribed constants  
 an  {\em interior-specification}   $\sint$: 
\begin{enumerate}[leftmargin=*]
 \item[-]
  Lower and upper bounds $\nint_\LB, \nint_\UB\in \mathbb{Z}_+$ 
  on   the number of interior-vertices 
of a target chemical graph~$\C$. 
  
\item[-] 
For each edge $e=u u'\in \Et\cup\Ew$, 
\begin{description} 
\item[]
 a lower bound $\ell_{\LB}(e)$ and 
 an upper bound $\ell_{\UB}(e)$  on the length $|E(P_e)|$ of
 a pure $u,u'$-path $P_e$. 
(For a notational convenience, set 
$\ell_\LB(e):=0$, $\ell_\UB(e):=1$, $e\in \Ez$ and
$\ell_\LB(e):=1$, $\ell_\UB(e):=1$, $e\in \Eew$.)
   
\item[]  
 a lower bound $\bl_{\LB}(e)$ and 
 an upper bound $\bl_{\UB}(e)$ on the number of leaf paths $Q_v$ attached 
 at  internal vertices $v$ of a pure $u,u'$-path $P_e$.   

\item[] 
 a lower bound $\ch_{\LB}(e)$ and 
 an upper bound $\ch_{\UB}(e)$  on the maximum 
 length  $|E(Q_v)|$ of a leaf path $Q_v$ attached  
 at an internal vertex $v\in V(P_e)\setminus\{u,u'\}$ 
 of a pure $u,u'$-path $P_e$.   
\end{description} 

\item[-]
For each vertex $v\in \VC$, 
\begin{description} 
\item[]
 a lower bound $\bl_{\LB}(v)$ and 
 an upper bound $\bl_{\UB}(v)$  on  
 the number of leaf paths $Q_v$ attached to $v$,
 where $0\leq \bl_{\LB}(v)\leq \bl_{\UB}(v)\leq 1$.
 
\item[]
 a lower bound $\ch_{\LB}(v)$ and 
 an upper bound $\ch_{\UB}(v)$  on the
 length $|E(Q_v)|$ of a leaf path $Q_v$ attached to $v$. 
\end{description}  

\item[-] 
For each edge $e=u u'\in \EC$, 
a lower bound $\bd_{m, \LB}(e)$ 
and an  upper bound $\bd_{m, \UB}(e)$  on
the number of edges with bond-multiplicity $m\in [2,3]$ in
$u,u'$-path $P_e$, where we regard $P_e$, $e  \in \Ez\cup \Eew$ 
as single edge $e$.
\end{enumerate}

We call a graph $H^*$ that satisfies an interior-specification $\sint$
a {\em $\sint$-extension of $\GC$}, 
where the bond-multiplicity of each edge has been determined.

Table~\ref{table:interior-spec}  shows an example of 
an interior-specification  $\sint$ to the seed graph  $\GC$ in 
Figure~\ref{fig:specification_example_1}. 

\begin{table}[h!]\caption{Example~1 of an interior-specification  $\sint$. }
\begin{tabular}{ |  c | c |  } \hline 
$\nint_\LB=20$ & $\nint_\UB = 28$ \\\hline 
\end{tabular}

 \begin{tabular}{ |  c | c c c c c c |  } \hline
                        & $a_1$ &  $a_2$ &   $a_3$ &   $a_4$ &   $a_5$ &   $a_6$   \\\hline
 $\ell_\LB(a_i)$&  2 &  2 &  2 & 3 &  2 &  1 \\ \hline
 $\ell_\UB(a_i)$&  3 & 4 &  3 & 5 & 4 &  4 \\\hline
 $\bl_\LB(a_i)$&  0 &  0 &   0 & 1 &  1 &   0 \\ \hline
 $\bl_\UB(a_i)$&  1 & 1 &   0 & 2 & 1 &   0 \\\hline
 $\ch_\LB(a_i)$&  0 &  1 & 0 & 4 &  3 &  0 \\ \hline
 $\ch_\UB(a_i)$&  3 & 3 &  1 & 6 & 5 &  2 \\\hline
\end{tabular} 

\begin{tabular}{ |  c | c c c c c c   c c c c  c c c |  } \hline
                        & $u_1$ &  $u_2$ &   $u_3$ &   $u_4$ &   $u_5$ &   $u_6$ 
                       & $u_7$ &   $u_8$ &   $u_9$ &   $u_{10}$ &   $u_{11}$ 
                       &   $u_{12}$ &   $u_{23}$ \\\hline 
 $\bl_\LB(u_i)$&  0 &  0 &   0 & 0 &  0 &   0
                       & 0 &   0 &  0 &   0 &  0 &  0 &  0 \\ \hline
 $\bl_\UB(u_i)$&  1 & 1 &   1 & 1 & 1 &   0
                       & 0 &   0 &  0 &   0 &  0 &  0 &  0\\\hline
 $\ch_\LB(u_i)$&  0 &  0 &   0 & 0 &  1 &   0
                       & 0 &   0 &  0 &   0 &  0 &  0 &  0 \\ \hline
 $\ch_\UB(u_i)$&  1 & 0 &   0 & 0 & 3 &   0
                       & 1 &   1 &  0 &   1 &  2 & 4 &  1 \\\hline
\end{tabular} 

\begin{tabular}{ |  c | c c c c c c   c c c c c c  c c c c c |  } \hline
                               & $a_1$ &  $a_2$ &   $a_3$ &   $a_4$ &   $a_5$ &   $a_6$ 
                               & $a_7$ &  $a_8$ &   $a_9$ &   $a_{10}$ &   $a_{11}$ &   $a_{12}$ 
                               & $a_{13}$ &   $a_{14}$ &   $a_{15}$ &   $a_{16}$ &   $a_{17}$  \\\hline
 $\bd_{2, \LB}(a_i)$ &  0    &  0 &   0 & 1 &  0 &   0
                                &  0   &  0 &  0 & 0 &  0 &   1
                                &  0    &  0 &   0 & 0     & 0  \\ \hline
 $ \bd_{2, \UB}(a_i)$&  1    & 1 &   0 & 2  & 2 &   0  
                                &  0    & 0&   0 & 0 &  0 &   1
                                &  0    &  0 &   0 & 0   & 0   \\ \hline
 $\bd_{3, \LB}(a_i)$ &  0    &  0 &   0 & 0 &  0 &   0
                                &  0   &  0 &  0 & 0 &  0 &   0
                                &  0    &  0 &   0 & 0   & 0   \\ \hline
 $ \bd_{3, \UB}(a_i)$&  0    & 0 &   0 & 0  & 1 &   0 
                                &  0    &  0 &   0 & 0 &  0 &   0
                                &  0    &  0 &   0 &  0    & 0   \\ \hline
\end{tabular} 
\label{table:interior-spec}  
\end{table}

\begin{figure}[h!] \begin{center}
\includegraphics[width=.58\columnwidth]{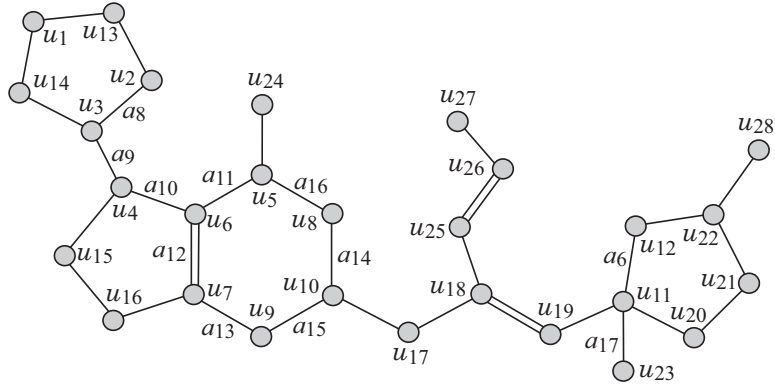}
\end{center} \caption{
An illustration of a graph 
$H^*$ that is obtained from  the seed graph  $\GC$ in 
Figure~\ref{fig:specification_example_1}
under the interior-specification  $\sint$ in 
Table~\ref{table:interior-spec}.    }
\label{fig:specification_example_3} \end{figure}

Figure~\ref{fig:specification_example_3} illustrates an example of 
an $\sint$-extension $H^*$ of seed graph  $\GC$ in 
Figure~\ref{fig:specification_example_1}
under the interior-specification  $\sint$ in 
Table~\ref{table:interior-spec}.  

\subsection*{Chemical-specification}
 
 Let $H^*$ be a graph that serves as 
 the interior $H^\inte$ of a target chemical graph $\C$,
 where the bond-multiplicity of each edge in $H^*$ has been determined.
 Finally we introduce a set of rules for constructing 
   a target chemical graph $\C$ from $H^*$ 
   by choosing  a chemical element $\ta\in \Lambda$ 
and assigning a ${\rho}$-fringe-tree $\psi$
 to each interior-vertex $v\in V^\inte$. 
We introduce the following rules for specifying
the size of $\C$, a set of chemical rooted trees  
that are allowed to use as  ${\rho}$-fringe-trees 
and lower and upper bounds on the frequency of
a chemical element, a chemical symbol, 
and an edge-configuration,
where we call the set of prescribed constants   
 a  {\em chemical specification} $\sce$:   
 
\begin{enumerate}[leftmargin=*]
\item[-] 
Lower and upper bounds $n_\LB,  n^*\in \mathbb{Z}_+$
on the number of vertices, where $\nint_\LB \leq n_\LB\leq n^*$.
 
\item[-] 
Subsets  $\mathcal{F}(v) \subseteq \mathcal{F}(D_\pi), v\in \VC$ 
and $\mathcal{F}_E \subseteq \mathcal{F}(D_\pi)$ 
 of chemical rooted trees $\psi$ with $\h(\anpsi)\leq {\rho}$, where 
 we require that 
 every ${\rho}$-fringe-tree $\C[v]$ rooted at a vertex $v\in \VC$
 (resp., at an internal vertex $v$ not in $\VC$)   in  $\C$ 
 belongs to $\mathcal{F}(v)$ (resp.,   $\mathcal{F}_E$).  
Let  $\mathcal{F}^*:=\mathcal{F}_E\cup \bigcup_{v\in \VC}\mathcal{F}(v)$
and 
$\Lambda^\ex$ denote the set of  chemical elements assigned to non-root
vertices over all chemical rooted trees in $\mathcal{F}^*$.  
 
\item[-] 
A subset  $\Lambda^\inte\subseteq \Lambda^\inte(D_\pi)$, where 
 we require that every chemical element $\alpha(v)$ 
 assigned to an interior-vertex  $v$ in $\C$ belongs to $\Lambda^\inte$.
Let $\Lambda:= \Lambda^\inte\cup \Lambda^\ex$ and
 $\na_\ta(\C)$ (resp., $\na_\ta^\inte(\C)$ and $\na_\ta^\ex(\C)$) 
 denote the number of vertices   (resp.,   interior-vertices and  exterior-vertices)
  $v$ such that $\alpha(v)=\ta$   in  $\C$.
 
\item[-] 
A set $\Ldg^\inte\subseteq \Lambda\times [1,4]$  of chemical  symbols
and  a set $\Gamma^\inte \subseteq \Gamma^\inte(D_\pi)$  
of  edge-configurations  $(\mu,\mu' ,m)$ with $\mu \leq \mu'$, where 
 we require that the edge-configuration $\ec(e)$ of an interior-edge $e$ in $\C$ 
 belongs to $\Gamma^\inte$.
We do not distinguish  $(\mu,\mu' ,m)$ and $(\mu' , \mu,m)$.

\item[-] 
Define  $\Gac^\inte$ to be the set of   adjacency-configurations such that  
$\Gac^\inte:=\{(\ta, \tb, m) \mid (\ta d, \tb d',m)\in \Gamma^\inte\}$.   
Let  $\ac_\nu^\inte(\C), \nu\in \Gac^\inte$   
denote  the number of  interior-edges $e$ such that $\ac(e)=\nu$  in $\C$.
  
\item[-] 
 Subsets $\Lambda^*(v)\subseteq \{\ta\in \Lambda^\inte\mid \val(\ta)\geq 2\}$, 
 $v\in \VC$,  
 we require that every chemical element $\alpha(v)$ 
 assigned to   a vertex $v\in  \VC$
 in the seed graph  belongs to $\Lambda^*(v)$.  

\item[-] Lower and upper bound functions 
$\na_\LB,\na_\UB: \Lambda\to  [1,n^*]$  and 
$\na_\LB^\inte,\na_\UB^\inte: \Lambda^\inte\to  [1,n^*]$ 
on the number of   interior-vertices  $v$ such that  $\alpha(v)=\ta$  in $\C$. 

\item[-] Lower and upper bound functions  
$\ns_\LB^\inte,\ns_\UB^\inte: \Ldg^\inte\to  [1,n^*]$ 
  on the number of   interior-vertices $v$ such that $\cs(v)=\mu$  in $\C$.   

\item[-] Lower and upper bound functions  
$\ac_\LB^\inte,\ac_\UB^\inte: \Gac^\inte \to  \mathbb{Z}_+$ 
 on the number of  interior-edges $e$ such that $\ac(e)=\nu$  in $\C$. 

\item[-] Lower and upper bound functions  
$\ec_\LB^\inte,\ec_\UB^\inte: \Gamma^\inte \to  \mathbb{Z}_+$ 
 on the number of  interior-edges $e$ such that $\ec(e)=\gamma$  in $\C$.  
 
 \item[-] Lower and upper bound functions  
$\fc_\LB,\fc_\UB: \mathcal{F}^*\to  [0,n^*]$ 
  on the number of   interior-vertices $v$ 
  such that $\C[v]$ is r-isomorphic to $\psi\in \mathcal{F}^*$  in $\C$.   
  
 \item[-] Lower and upper bound functions  
$\ac^\lf_\LB,\ac^\lf_\UB: \Gac^\lf \to  [0,n^*]$ 
  on the number of  leaf-edges $uv$ in $\acC$
  with adjacency-configuration $\nu$.  
\end{enumerate}
 
We call a chemical graph $\C$ that satisfies a chemical specification $\sce$
a {\em $(\sint,\sce)$-extension of $\GC$},
and denote by $\mathcal{G}(\GC, \sint,\sce)$ the set of
all $(\sint,\sce)$-extensions of $\GC$. 

Table~\ref{table:chemical_spec}  shows an example of 
a chemical-specification  $\sce$ to the seed graph  $\GC$
 in Figure~\ref{fig:specification_example_1}. 
 

\begin{table}[h!]\caption{Example~2 of a chemical-specification  $\sce$.  
}
\begin{tabular}{ |  l |  } \hline
 $n_\LB=30$,  $n^* =50$. \\\hline
  branch-parameter:   ${\rho}=2$  \\\hline
\end{tabular}

\begin{tabular}{ |  l |  } \hline
 Each of sets $\mathcal{F}(v), v\in \VC$ and
 $\mathcal{F}_E$ is set to be \\
 the set $\mathcal{F}$  of chemical rooted trees $\psi$ with $\h(\anpsi)\leq {\rho}=2$
in Figure~\ref{fig:specification_example_1}(b). \\\hline
\end{tabular}

\begin{tabular}{ |  c | c |   } \hline
  $\Lambda=\{ \ttH,\ttC,\ttN,\ttO, \ttS_{(2)},\ttS_{(6)}, \ttP=\ttP_{(5)}\}$ & 
  $\Ldg^\inte =\{ \ttC2 , \ttC3,  \ttC4, \ttN2, \ttN3, \ttO2,
    \ttS_{(2)}2,  \ttS_{(6)}3, \ttP4   \}$  
\\\hline
\end{tabular}

\begin{tabular}{ |  c | l |  } \hline
  $\Gac^{\inte}$ &
  $ \nu_1 \!=\!(\ttC   , \ttC  , 1) ,   \nu_2 \!=\!(\ttC   , \ttC  , 2) ,   
   \nu_3 \!=\!(\ttC   , \ttN  , 1) ,  \nu_4 \!=\!(\ttC  , \ttO  , 1), 
    \nu_5 \!=\! (\ttC, \ttS_{(2)}, 1),\nu_6 \!=\!(\ttC  , \ttS_{(6)}, 1), 
    \nu_7 \!=\! (\ttC  , \ttP  , 1) $  \\ \hline
\end{tabular}

\begin{tabular}{ |  c | l |  } \hline
  $\Gamma^{\inte}$ &
  $ \gamma_1 \!=\! (\ttC 2 , \ttC 2, 1) ,
   \gamma_2 \!=\!(\ttC 2 , \ttC 3, 1) ,  
   \gamma_3 \!=\!(\ttC 2 , \ttC 3, 2) ,  
   \gamma_4 \!=\!(\ttC 2 , \ttC 4, 1) , 
   \gamma_5 \!=\!(\ttC 3 , \ttC 3, 1) , 
   \gamma_6 \!=\!(\ttC 3 , \ttC 3, 2) , $ \\
   &
  $   
    \gamma_7 \!=\!(\ttC 3 , \ttC 4, 1), 
   \gamma_8 \!=\!(\ttC 2 , \ttN 2, 1) ,  
   \gamma_9 \!=\!(\ttC 3 , \ttN 2, 1) ,  
   \gamma_{10} \!=\!(\ttC 3 , \ttO 2, 1), 
    \gamma_{11} \!=\!(\ttC 2 , \ttC 2, 2),  
    \gamma_{12} \!=\!(\ttC 2 , \ttO 2, 1) ,$ \\
   &
  $  
    \gamma_{13} \!=\!(\ttC 3 , \ttN3, 1), 
    \gamma_{14} \!=\!(\ttC 4, \ttS_{(2)} 2, 2),  
    \gamma_{15} \!=\!(\ttC 2 , \ttS_{(6)}3, 1), 
   \gamma_{16} \!=\!(\ttC 3 , \ttS_{\tiny (6)}3, 1), 
    \gamma_{17} \!=\!(\ttC 2, \ttP4, 2), $ \\
   &
  $  
    \gamma_{18} \!=\!(\ttC 3, \ttP4, 1)  
     $ \\ \hline
\end{tabular}
    
\begin{tabular}{ |  l|  } \hline
$\Lambda^*(u_1)=\Lambda^*(u_8)=\{{\tt C,  N}\}$, 
$\Lambda^*(u_9)=\{{\tt C, O}\}$, 
   $\Lambda^*(u)=\{\ttC\}$, $u\in \VC\setminus\{u_1,u_8,u_9\}$
   \\\hline
\end{tabular}

\begin{tabular}{ |  c | c c c c  c c c |  } \hline
                         & ${\tt H}$  & ${\tt C}$ &   ${\tt N}$ &     ${\tt O}$ 
                         & $\ttS_{(2)}$ & $\ttS_{(6)}$ & $\ttP$  \\\hline
 $\na_\LB(\ta)$ & 40 &  27 &  1 &   1 & 0 & 0 & 0   \\ \hline 
 $\na_\UB(\ta)$ & 65 & 37 & 4 &  8  &   1 &   1 &   1 \\\hline
\end{tabular} 
\begin{tabular}{ |  c | c c c  c c c   |  } \hline
   & $\ttC$ &   $\ttN$ &     $\ttO$  & $\ttS_{(2)}$ & $\ttS_{(6)}$ & $\ttP$  \\\hline
 $\na_\LB^{\inte}(\ta)$ &   9 &  1 &   0  & 0 & 0 & 0      \\ \hline
 $\na_\UB^{\inte}(\ta) $&  23 & 4 & 5 &   1 &   1 &   1  \\\hline
\end{tabular} 

\begin{tabular}{ |  c | c c c c c c  c c c   |  } \hline
    & $\ttC2$ &  $\ttC3$ &   $\ttC4$ & $\ttN2$ &   $\ttN3$ &   $\ttO2$
   & $\ttS_{(2)}2$ & $\ttS_{(6)}3$ & $\ttP4$  \\\hline
 $\ns_\LB^{\inte}(\mu)$ &  3 &  5 &   0 & 0 &  0 &   0 & 0 &  0 &   0    \\ \hline
 $\ns_\UB^{\inte}(\mu) $&  8 & 15 & 2 & 2 & 3 &  5  &   1 &   1 &   1   \\\hline
\end{tabular} 

\begin{tabular}{ |  c | c c c c c c c |  } \hline
         & $\nu_1 $ &   $\nu_2 $ & $\nu_3 $   & $\nu_4 $
         &   $\nu_5 $ & $\nu_6 $   & $\nu_7 $ \\\hline
 $\ac_\LB^{\inte}(\nu)$  &  0  &  0  & 0  & 0   & 0  & 0 & 0     \\ \hline
 $\ac_\UB^{\inte}(\nu)$ &  30 & 10 & 10 & 10 & 1 & 1 & 1 \\\hline
\end{tabular} 

\begin{tabular}{ |  c | c c c c c c c c c c c c c c c c c c |  } \hline
    & $\gamma_1 $ &   $\gamma_2 $ & $\gamma_3 $   & $\gamma_4 $ 
     & $\gamma_5 $
    & $\gamma_6 $ &   $\gamma_7 $ & $\gamma_8 $   & $\gamma_9 $ 
     & $\gamma_{10} $   & $\gamma_{11} $       & $\gamma_{12} $          
     & $\gamma_{13} $   & $\gamma_{14} $       & $\gamma_{15} $          
     & $\gamma_{16} $   & $\gamma_{17} $       & $\gamma_{18} $          
                            \\\hline
 $\ec_\LB^{\inte}(\gamma)$ &  0 &  0 & 0 &  0  & 0 &  0 &  0 & 0 &  0  & 0 & 0 & 0 
    &  0 & 0 &  0  & 0 & 0 & 0  \\ \hline
 $\ec_\UB^{\inte}(\gamma) $& 4 & 15 & 4 &  4  & 10 &  5 & 4 & 4 &  6 & 4 & 4 & 4
 &  2 & 2 &  2  & 2 & 2 & 2  \\\hline
\end{tabular}

\begin{tabular}{ |  c | c   c   |  } \hline 
& $\psi\in\{\psi_i\mid i=1,6,11\}$ 
& $\psi\in \mathcal{F}^*\setminus \{\psi_i\mid i=1,6,11\}$ \\\hline
 $\fc_\LB(\psi)$  &  1 &    0   \\ \hline 
 $\fc_\UB(\psi)$ &  10 &  3\\\hline
\end{tabular}

\begin{tabular}{ |  c | c   c   |  } \hline 
& $\nu\in\{(\ttC,\ttC,1),(\ttC,\ttC,2)  \}$ 
& $\nu\in \Gac^\lf \setminus \{(\ttC,\ttC,1),(\ttC,\ttC,2)  \}$   \\\hline
 $\ac^\lf_\LB(\nu)$  &  0 &    0   \\ \hline 
 $\ac^\lf_\UB(\nu)$ &  10 &  8 \\\hline
\end{tabular} 

\label{table:chemical_spec}
\end{table}

Figure~\ref{fig:example_chemical_graph} 
 illustrates an example $\Co$ of 
a   $(\sint,\sce)$-extension of $\GC$   obtained 
from the  $\sint$-extension $H^*$  
 in Figure~\ref{fig:specification_example_3} 
under the chemical-specification $\sce$ in Table~\ref{table:chemical_spec}.  


\section{Test Instances for Inferring Chemical Graphs}\label{sec:test_instances} 

We prepared the following instances (a)-(d) for conducting experiments
of the second phase of the framework. 
 
 In the second phase of inferring chemical graphs, we  use two properties 
 $\pi\in \{${\sc At}, {\sc FlmL}$\}$ 
 and define a set $\Lambda(\pi)$ of chemical elements as follows:  
 $\Lambda(${\sc At}$)=\Lambda_3=\{\ttH,\ttC,\ttO, \ttN, \ttCl, \ttS_{(2)}, \ttS_{(6)} \}$
 and 
 $\Lambda(${\sc FlmL}$)=\Lambda_6=\{\ttH,\ttC,\ttO, \ttN, \ttCl ,\ttP_{(2)},\ttP_{(5)} \}$. 
 
\begin{itemize} 
  \item[(a)]  $I_{\mathrm{a}} =(\GC,\sint,\sce)$: The instance
  introduced in Section~\ref{sec:specification} to explain the target specification.
For each property $\pi$, we replace
 $\Lambda=\{ \ttH,\ttC,\ttO, \ttN, \ttS_{(2)},\ttS_{(6)},\ttP_{(5)}\}$
in Table~\ref{table:chemical_spec} 
 with $\Lambda(\pi)
 \cap \{\ttH,\ttC,\ttO, \ttN, \ttS_{(2)},\ttS_{(6)},\ttP_{(5)}\}$
 and  remove from the $\sce$
 all chemical symbols,  edge-configurations and fringe-configurations
  that cannot be constructed from the replaced element set 
 (i.e., those containing a chemical element  $ \ttP_{(5)}$). 
 \end{itemize}

\begin{itemize} 
  \item[(b)] $I_\mathrm{b}^i=(\GC^i,\sint^i, \sce^i)$, $i=1,2,3,4$:
 An instance for inferring chemical graphs with rank at most 2.  
In the four instances $I_\mathrm{b}^i$, $i=1,2,3,4$, 
the following specifications in $(\sint,\sce)$ are common. 
\begin{enumerate}
\item[] 
Set  $\Lambda:=\Lambda(\pi)$
 for a given property $\pi\in \{${\sc  At,  FlmL}$\}$, 
 set $\Ldg^\inte$ to be
the set of all possible symbols in $\Lambda\times[1,4]$  
that appear in the data set $\mathcal{C}_\pi$  
and set $\Gamma^\inte$
to be the set  of 
 all  edge-configurations that appear in the data set $\mathcal{C}_\pi$. 
Set  $\Lambda^*(v):= \Lambda$,  $v\in \VC$. 
 
\item[] 
The lower bounds  
 $\ell_\LB $, $\bl_\LB $, $\ch_\LB $,  
 $\bd_{2,\LB}$,   $\bd_{3,\LB}$,  
 $\na_\LB$,  $\na^\inte_\LB$,  $\ns^\inte_\LB$,  
$\ac^\inte_\LB$, $\ec^\inte_\LB$ and $\ac^\lf_\LB$  are all set to be 0.

\item[] 
Set  upper bounds   
 $\na_\UB(\ta):=n^*, \na\in\{\ttH,\ttC\}$,   
 $\na_\UB(\ta):=5, \na\in\{\ttO,\ttN\}$,
 $\na_\UB(\ta):=2, \na\in\Lambda\setminus \{\ttH,\ttC,\ttO,\ttN\}$. 
The other upper bounds  
 $\ell_\UB $, $\bl_\UB $, $\ch_\UB $,  
 $\bd_{2,\UB}$,   $\bd_{3,\UB}$,  
 $\na^\inte_\UB$,  $\ns^\inte_\UB$,  
$\ac^\inte_\UB$, $\ec^\inte_\UB$ and $\ac^\lf_\UB$ 
are all set to be an upper bound $n^*$  on $n(G^*)$.

\item[] 
We specify $n_\LB$ as a parameter and
set
$n^*:=n_\LB+10$,
  $\nint_\LB:=\lfloor (1/4) n_\LB \rfloor$ and
   $\nint_\UB:=\lfloor (3/4) n_\LB \rfloor$. 

\item[] 
   For each property $\pi$, let $\mathcal{F}(\mathcal{C}_\pi)$ denote
    the set of 2-fringe-trees in the compounds in $\mathcal{C}_\pi$,
   and select a subset $\mathcal{F}_\pi^i\subseteq  \mathcal{F}(\mathcal{C}_\pi)$ with
   $|\mathcal{F}_\pi^i|=45-5i$, $i\in [1,5]$.
   For each instance $I_\mathrm{b}^i$, 
   set $\mathcal{F}_E :=\mathcal{F}(v):=  \mathcal{F}_\pi^i$, $v\in \VC$ and 
$\fc_\LB(\psi):=0, \fc_\UB(\psi):=10, \psi\in  \mathcal{F}_\pi^i$. 
\end{enumerate}
 
  Instance $I_\mathrm{b}^1$ is given   by the rank-1 seed graph $\GC^1$ 
  in Figure~\ref{fig:specification_example_polymer}(i)
  and   Instances $I_\mathrm{b}^i$, $i=2,3,4$ are
   given by  the rank-2 seed graph $\GC^i$, $i=2,3,4$ in 
   Figure~\ref{fig:specification_example_polymer}(ii)-(iv).

\begin{itemize} 
 \item[(i)]  For instance $I_\mathrm{b}^1$, select as a seed graph 
  the monocyclic graph   $\GC^1=(\VC,\EC=\Et\cup \Ew)$
  in Figure~\ref{fig:specification_example_polymer}(i),
  where $\VC=\{u_1,u_2\}$, $\Et=\{a_1\}$ and  $ \Ew=\{a_2\}$. 
We  include a linear constraint 
$\ell(a_1)\leq \ell(a_2)$ 
and $5\leq \ell(a_1)+\ell(a_2) \leq 15$  as part of the side constraint. 
  
 \item[(ii)]
 For instance $I_\mathrm{b}^2$, select as a seed graph 
  the  graph   $\GC^2=(\VC,\EC=\Et\cup \Ew\cup \Eew)$ 
  in Figure~\ref{fig:specification_example_polymer}(ii),
  where
$\VC=\{u_1,u_2,u_3,u_4\}$, 
$\Et=\{a_1,a_2\}$, 
$\Ew=\{a_3\}$  and 
$\Eew=\{a_4,a_5\}$. 
%
We include a linear constraint $\ell(a_1)\leq \ell(a_2)$ 
and $\ell(a_1)+\ell(a_2)+\ell(a_3)\leq 15$. 
    
 \item[(iii)]
 For instance $I_\mathrm{b}^3$, select as a seed graph 
  the  graph   $\GC^3=(\VC,\EC=\Et\cup \Ew\cup \Eew)$ 
  in Figure~\ref{fig:specification_example_polymer}(iii),   where
$\VC=\{u_1,u_2,u_3,u_4\}$, 
$\Et=\{a_1\}$, 
$\Ew=\{a_2, a_3\}$  and 
$\Eew=\{a_4,a_5\}$. 
%
We include   linear constraints 
$\ell(a_1)\leq \ell(a_2)+\ell(a_3)$, $\ell(a_2)\leq \ell(a_3)$
and $\ell(a_1)+\ell(a_2)+\ell(a_3)\leq 15$.  

 \item[(iv)] 
 For instance $I_\mathrm{b}^4$, select as a seed graph 
  the  graph   $\GC^4=(\VC,\EC=\Et\cup \Ew\cup \Eew)$ 
  in Figure~\ref{fig:specification_example_polymer}(iv),   where
$\VC=\{u_1,u_2,u_3,u_4\}$, 
$\Ew=\{a_1, a_2, a_3\}$  and 
$\Eew=\{a_4,a_5\}$. 
We   include   linear constraints 
$\ell(a_2)\leq \ell(a_1)+1$,
$\ell(a_2)\leq \ell(a_3)+1$,  $\ell(a_1)\leq \ell(a_3)$  
and $\ell(a_1)+\ell(a_2)+\ell(a_3)\leq 15$. 
 \end{itemize}
 \end{itemize}

 We define instances in (c) and (d) 
 in order to find chemical graphs that have an intermediate structure of
 given two chemical cyclic graphs 
 $G_A=(H_A=(V_A,E_A),\alpha_A,\beta_A)$ 
and $G_B=(H_B=(V_B,E_B),\alpha_B,\beta_B)$.
Let
 $\Lambda_A^\inte$ and  $\Lambda_{\mathrm{dg},A}^\inte$ 
 denote the sets  of chemical elements
 and chemical symbols  of
 the interior-vertices in $G_A$, 
 $\Gamma_A^\inte$   denote the sets of edge-configurations of
  the interior-edges in $G_A$,   
  and 
  $\mathcal{F}_A$ denote the set of 2-fringe-trees in $G_A$.  
Analogously define sets
 $\Lambda_B^\inte$,    $\Lambda_{\mathrm{dg},B}^\inte$,   
 $\Gamma_B^\inte$ and   $\mathcal{F}_B$ 
 in $G_B$.

\begin{itemize}  
\item[(c)]  $I_{\mathrm{c}}=(\GC,\sint,\sce)$: 
An instance aimed to infer a chemical graph $G^\dagger$ such that
the core of $G^\dagger$ is equal to the core of $G_A$ and 
the frequency of each edge-configuration in the non-core of $G^\dagger$
is equal to that of  $G_B$. 
We use chemical compounds CID~24822711 and CID~59170444 in 
 Figure~\ref{fig:instance_I_c_I_d}(a) and (b)
 for $G_A$ and $G_B$, respectively.  \\
Set   a seed graph $\GC=(\VC,\EC=\Eew)$ to be the core of $G_A$. \\
Set  $\Lambda:=\{{\tt H,C,N,O}\}$, 
and  set $\Ldg^\inte$ to be
the set of all possible chemical symbols in $\Lambda\times[1,4]$.\\
Set 
$\Gamma^\inte:=\Gamma_A^\inte\cup \Gamma_B^\inte$ and 
  $\Lambda^*(v):=\{\alpha_A(v)\}$, $v\in \VC$.  \\
Set 
$\nint_\LB:=\min\{\nint(G_A), \nint(G_B)\}$, 
$\nint_\UB:=\max\{\nint(G_A), \nint(G_B)\}$, \\
$n_\LB:=\min\{n(G_A), n(G_B)\}-10=40$ 
and   $n^*:=\max\{n(G_A), n(G_B)\}+5$. \\
Set  lower bounds  
 $\ell_\LB $, $\bl_\LB $, $\ch_\LB $,  
 $\bd_{2,\LB}$,   $\bd_{3,\LB}$,  
 $\na_\LB$,  $\na^\inte_\LB$,  $\ns^\inte_\LB$, 
$\ac^\inte_\LB$ and  $\ac^\lf_\LB$  to be 0.\\
Set  upper bounds   
 $\na_\UB(\ta):=n^*, \na\in\{\ttH,\ttC\}$,   
 $\na_\UB(\ta):=5, \na\in\{\ttO,\ttN\}$,
 $\na_\UB(\ta):=2, \na\in\Lambda\setminus \{\ttH,\ttC,\ttO,\ttN\}$ 
and set the other upper bounds
 $\ell_\UB $, $\bl_\UB $, $\ch_\UB $,  
 $\bd_{2,\UB}$,   $\bd_{3,\UB}$,  
$\na^\inte_\UB$,  $\ns^\inte_\UB$, 
$\ac^\inte_\UB$   and  $\ac^\lf_\UB$ to be  $n^*$. \\
Set $\ec_\LB^\inte(\gamma)$ 
to be the number of core-edges  in $G_A$ with $\gamma\in \Gamma^\inte$ and  
 $\ec_\UB^\inte(\gamma)$  
to be the number interior-edges in $G_A$ and  $G_B$ 
with edge-configuration $\gamma$. \\
Let $\mathcal{F}_B^{(p)}, p\in [1,2]$ denote the set of chemical rooted 
trees r-isomorphic $p$-fringe-trees in $G_B$; \\
Set $\mathcal{F}_E :=\mathcal{F}(v):= 
 \mathcal{F}_B^{(1)}\cup \mathcal{F}_B^{(2)}$, $v\in \VC$ and
$\fc_\LB(\psi):=0, \fc_\UB(\psi):=10, \psi\in \mathcal{F}_B^{(1)}\cup \mathcal{F}_B^{(2)}$. 
 
  \item[(d)] $I_{\mathrm{d}}=(\GC^1,\sint, \sce)$:     
An instance aimed to infer a chemical monocyclic graph $G^\dagger$ such that
the frequency vector of  edge-configurations in  $G^\dagger$
is a vector obtained by merging those of $G_A$ and $G_B$.
We use chemical monocyclic compounds CID~10076784 and CID~44340250
in   Figure~\ref{fig:instance_I_c_I_d}(c) and (d) 
 for $G_A$ and $G_B$, respectively.  
Set a seed graph to be   the monocyclic seed graph  
 $\GC^1=(\VC,\EC=\Et\cup \Ew)$ with 
  $\VC=\{u_1,u_2\}$, $\Et=\{a_1\}$ and  $ \Ew=\{a_2\}$ 
  in Figure~\ref{fig:specification_example_polymer}(i). \\
Set  $\Lambda:=\{{\tt H,C,N,O}\}$,  
 $\Ldg^\inte:=\Lambda_{\mathrm{dg},A}^\inte 
                 \cup \Lambda_{\mathrm{dg},B}^\inte$ and 
$\Gamma^\inte:=\Gamma_A^\inte\cup \Gamma_B^\inte$. \\
Set 
$\nint_\LB:=\min\{\nint(G_A), \nint(G_B)\}$, 
$\nint_\UB:=\max\{\nint(G_A), \nint(G_B)\}$, \\
  $n_\LB:=\min\{n(G_A),n(G_B)\}=40$ and  
  $n^*:=\max\{n(G_A),n(G_B)\}$. \\
Set  lower bounds  
 $\ell_\LB $, $\bl_\LB $, $\ch_\LB $,  
 $\bd_{2,\LB}$,   $\bd_{3,\LB}$,  
 $\na_\LB$,  $\na^\inte_\LB$,  $\ns^\inte_\LB$, 
$\ac^\inte_\LB$  and  $\ac^\lf_\LB$ to be 0.\\
Set  upper bounds   
 $\na_\UB(\ta):=n^*, \na\in\{\ttH,\ttC\}$,   
 $\na_\UB(\ta):=5, \na\in\{\ttO,\ttN\}$,
 $\na_\UB(\ta):=2, \na\in\Lambda\setminus \{\ttH,\ttC,\ttO,\ttN\}$ 
and set the other  upper bounds  
 $\ell_\UB $, $\bl_\UB $, $\ch_\UB $,  
 $\bd_{2,\UB}$,   $\bd_{3,\UB}$,  
  $\na^\inte_\UB$,  $\ns^\inte_\UB$,
$\ac^\inte_\UB$ and  $\ac^\lf_\UB$  to be   $n^*$. \\
For each edge-configuration
 $\gamma \in \Gamma^\inte$,  
let  $\x^*_A(\gamma^\inte)$  (resp., $\x^*_B(\gamma^\inte)$)   denote
 the number of interior-edges with $\gamma$ in $G_A$ (resp., $G_B$), 
 $\gamma \in \Gamma^\inte$ and   
set \\
$\x^*_{\min}(\gamma):=\min\{\x^*_A(\gamma), \x^*_B(\gamma)\}$, 
 $\x^*_{\max}(\gamma):=\max\{\x^*_A(\gamma), \x^*_B(\gamma)\}$, \\
$\ec_\LB^\inte(\gamma):=
\lfloor (3/4)\x^*_{\min}(\gamma)+(1/4)\x^*_{\max}(\gamma) \rfloor$
and  \\
$\ec_\UB^\inte(\gamma):=
\lceil (1/4)\x^*_{\min}(\gamma)+(3/4)\x^*_{\max}(\gamma) \rceil$. \\
Set $\mathcal{F}_E :=\mathcal{F}(v):=  \mathcal{F}_A\cup \mathcal{F}_B$, 
$v\in \VC$ and 
$\fc_\LB(\psi):=0, \fc_\UB(\psi):=10, \psi\in \mathcal{F}_A\cup \mathcal{F}_B$. \\
We  include a linear constraint 
$\ell(a_1)\leq \ell(a_2)$ 
and $5\leq \ell(a_1)+\ell(a_2) \leq 15$  as part of the side constraint. 
 \end{itemize}

\end{document}